\documentclass[twocolumn,abs,prb,showpacs]{revtex4-1}

\usepackage{graphicx}
\usepackage{dcolumn}
\usepackage{float}
\usepackage{amsmath}
\usepackage{bm}

\newcommand{\comment}[1]{}

\begin{document}


\title{Dynamical conductivity of AA-stacked bilayer graphene}

\author{C.J. Tabert$^{1,2}$}
\author{E.J. Nicol$^{1,2,3}$}
\affiliation{$^1$Department of Physics, University of Guelph,
Guelph, Ontario N1G 2W1 Canada} 
\affiliation{$^2$Guelph-Waterloo Physics Institute, University of Guelph, Guelph, Ontario N1G 2W1 Canada}
\affiliation{$^3$Kavli Institute for Theoretical Physics, University of California, Santa Barbara, CA 93106 USA}
\date{\today}

\begin{abstract}
{We calculate the dynamical
conductivity of AA-stacked bilayer graphene as a function
of frequency and in the presence of a finite chemical potential due to charging.
Unlike the monolayer, we find a Drude absorption at charge neutrality in addition
to an interband absorption with onset of twice the interlayer hopping energy.
At finite doping, the interband absorption exhibits two edges 
which depend on both chemical potential and
interlayer hopping energy. We study the behaviour as a function of
varying chemical potential
 relative to the interlayer hopping energy scale and compute the 
partial optical sum. The results are contrasted with the previously
published case of AB-stacking.
While we focus on in-plane conductivity, we also 
provide the perpendicular conductivity for both AB and AA stacking.
We also examine conductivity for other variations with AA-stacking, 
such as AAA-stacked trilayer.
Based on proposed models for topological insulators discussed in the
literature, we also consider the effect of spin orbit coupling on the
optical properties of an AA-stacked bilayer which illustrates the effect
of an energy gap opening at  points in the band structure.
}
\end{abstract}


\maketitle

\section{Introduction}
Graphene continues to provide a rich platform for investigations
into the physics of massless Dirac fermions. Initially studies of graphene 
were limited to the realm of theory where the low energy linear 
dispersion\cite{Wallace:1947} and chiral nature of the honeycomb carbon lattice 
were shown\cite{Semenoff:1984} to result from a simple nearest-neighbor-hopping
 tight-binding Hamiltonian which at low energy maps on to a Dirac Hamiltonian
for massless fermions with Fermi velocity $v_F$. With the experimental 
realization of graphene\cite{Novoselov:2004,Novoselov:2005a}, 
a considerable literature has now accumulated which has uncovered a variety of exotic effects, 
such as an unusual quantum Hall effect\cite{Zhang:2005,Novoselov:2005b}, 
giant Faraday rotation\cite{Crassee:2011}, plasmarons\cite{Bostwick:2010}, and so on, some 
of which has been summarized in reviews\cite{Neto:2009, Abergel:2010, DasSarma:2011, Kotov:2010}. 

\begin{figure}[h!]
\includegraphics[width=0.9\linewidth]{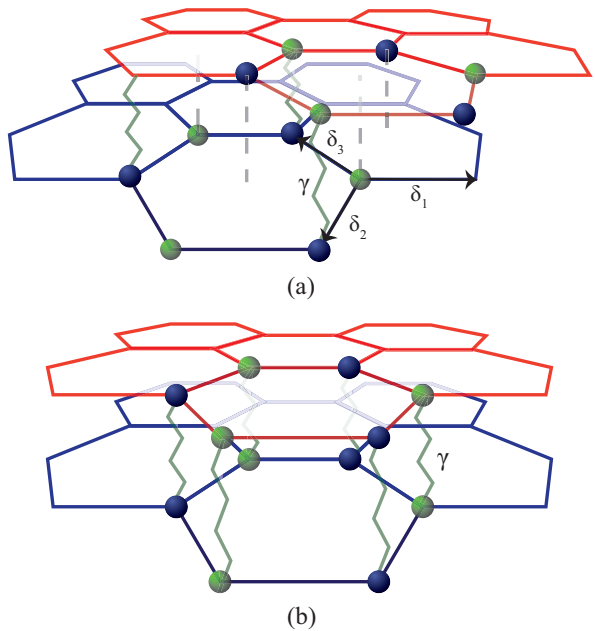}
\caption{(Color online)
(a) Two AB-stacked graphene sheets with the dark (blue) dots representing
one sublattice and  the light (green) dots, the other. (b) Two AA-stacked graphene sheets.
}\label{fig:Stacking}
\end{figure}

Bilayer graphene
is also of intense interest as it too shows an unusual quantum Hall effect\cite{Geim:2007,McCann:2006a}
and indeed its low energy tight-binding Hamiltonian maps to an equation
for chiral fermions with an effective mass\cite{McCann:2012} based on an interlayer hopping
parameter $\gamma$. In addition, it has been seen that bilayer graphene can
develop a sizeable band gap which is tunable by charge doping.\cite{Ohta:2006} 
Recent interest in bilayer graphene physics has focused on the large degeneracy
at the charge neutrality point which provides opportunity for instabilities
leading to new ground states. See Ref.~\cite{McCann:2012} for a summary of
the literature on this point and also a general review of the
properties of bilayer graphene. The natural form for bilayer 
graphene is the so-called Bernal or AB-stacking which is the basis of
the parent compound graphite from which it is usually derived. 
Consequently,
past work has primarily focused on this stacking configuration. However,
more recently, Moir\'e patterns seen in scanning tunneling microscopy imaging
of graphene bilayers and multilayers point to alternative stackings where
one layer is rotated by some angle relative to the other.\cite{Li:2010,Luican:2011} 
This is sometimes
referred to as twisted or misaligned bilayer graphene. In these systems,
the electronic properties are modified at low energy such that
monolayer behaviour appears along with a reduced Fermi velocity.\cite{Luican:2011} In
these systems, it is possible to have regions which are rich in AB-stacking
and regions which display mainly AA-stacking. These types of stackings are shown in 
Fig.~\ref{fig:Stacking}(a) and (b), respectively. Here, A and B refer to atoms on the 
two triangular sublattices of the honeycomb lattice and the stacking is in  reference to
whether the A atom of one plane is stacked over the A or B atom of the other
plane. For Bernal AB-stacking only half the atoms are aligned on top of each other 
and the other half sit over the center of the hexagon in the opposite layer.
For AA-stacking all atoms are matched up between the two layers.
For very small twist angles, the regions of AA-stacking have been suggested to
provide localization effects.\cite{Trambly:2010, Trambly:arxiv} Very recently, AA-stacked 
graphene has also attracted interest due to research which has identified
such stacking in certain samples, potentially making this another
experimentally accessible system to study.\cite{Liu:2009,Borysiuk:2011} 
For AA-stacking there is also the prediction for new ground states to occur,
such as antiferromagnetism\cite{Rakhmanov:2011}.

As a result, we are motivated by these developments to examine the dynamical
conductivity of doped AA-stacked graphene to elucidate features which would
demonstrate unique properties of this system and allow for the identification
of characteristic energy scales associated with the band structure. Moreover,
as the optical properties of graphene are of considerable importance for 
technological applications, all variants of graphene are also of potential
interest and should be examined. The dynamical conductivity of graphene
has been extensively studied 
theoretically\cite{Ando:2002,Gusynin:2006a,Gusynin:2006,Falkovsky:2007,Peres:2006,Gusynin:2007} 
and experiments have largely verified the expected behaviour\cite{Li:2008,Kuzmenko:2008,Nair:2008,Mak:2008}. 
Likewise, the conductivity for Bernal-stacked bilayer graphene has been
predicted theoretically\cite{Abergel:2007,Nicol:2008,Nilsson:2008,Koshino:2009} and observed\cite{Li:2009,Zhang:2008,Kuzmenko:2009}. 
There has also been work on magneto-optical conductivity of graphene in which
theory and experiment are also in good agreement. Indeed, a review of this
literature may be found in Ref.~\cite{Orlita:2010}. Some preliminary work on the
absorption coefficient of undoped AA-stacked graphene in zero magnetic
field has been reported\cite{Xu:2010}
however most materials naturally occur with charge doping where the Fermi level or
chemical potential $\mu$ is away from charge neutrality ($\mu=0$). Furthermore, the 
interesting feature for practical applications is the variation of optical
properties with doping, usually achieved through a field effect transistor structure.

In the following, we provide a thorough examination of the finite frequency
 conductivity for AA-stacking, 
for both in-plane and out-of-plane response.
Unlike a single monolayer, AA-stacked graphene shows a Drude response
in the in-plane conductivity at charge neutrality along with Pauli blocking
at low frequencies below the onset of a flat interband absorption. This interband
absorption splits at finite doping into  two interband absorption
edges leading to flat universal values associated with one and two layers. In
terms of the interlayer hopping energy $\gamma$, the partial optical sum, which is
a measure of the transfer of spectral weight, shows
distinct behaviour for $\mu<\gamma$ versus $\mu>\gamma$. The perpendicular
conductivity has a strong response at $2\gamma$ at all dopings.  This is contrasted
with the case for AB-stacking which we also show here as we can also provide
an analytical formula for this quantity to add to the literature. 
Indeed, we have provided analytical formulae
in almost all cases and physical understanding of our results are given. 
We also 
contrast the AA-stacked case with that for AAA-stacking. Finally, because of
the connection between the Dirac nature of graphene and topological insulators (TIs),
we also  follow-up on a toy-model\cite{Prada:2011} by providing the
conductivity for two AA-stacked sheets
with spin orbit coupling in one or both planes, potentially mimicking weakly coupled
TIs or a TI in proximity to a metallic sheet.

Our paper is organized as follows. In Section~II, we review our theoretical calculation
for the dynamical conductivity of AA-stacked graphene. Our presentation follows that for AB-stacked bilayer graphene done by Nicol and Carbotte\cite{Nicol:2008} which uses
many-body Green's function which easily allows for further theoretical development, such
as the inclusion of a self-energy from impurities\cite{Nilsson:2008}, electron-phonon 
interaction\cite{Stauber:2008a,Nicol:2009,Carbotte:2010,Pound:2011a,Pound:2011b,Pound:2012}, electron-electron
interactions\cite{LeBlanc:2011,Carbotte:2012,Principi:2012}, etc. 
In Section~III, we discuss the results of the AA-stacked
case, examining both in-plane and perpendicular conductivity and contrasting with the AB-stacked
case. We discuss the effect of biasing the bilayer.
We also consider the theory for other variations on the AA-stacked case in the
subsequent sections.
For instance, we examine the case for the AAA-stacked trilayer in Section~IV and report results for
models with spin orbit coupling in AA-stacked bilayer in Section~V. Our conclusions are found in Section~VI.
  
\section{Theory for AA-stacked bilayer}

To derive the optical conductivity of AA-stacked bilayer graphene, we follow the method shown in 
the work of Nicol and Carbotte\cite{Nicol:2008} for the case of AB-stacked bilayer graphene. This is
based on the Kubo formula for the current-current response function
and the many-body Green's function approach.\cite{Mahan:1990} Thus, to begin 
we must first examine the band structure and provide an expression for the electronic Green's function.  
For the case of AA stacking, an A (B) atom in the upper layer is stacked directly above an A (B) atom in the lower layer, 
see Fig. \ref{fig:Stacking}(b), as opposed to the typical Bernal stacking shown in Fig. \ref{fig:Stacking}(a).

\begin{figure}
\includegraphics[width=1.0\linewidth]{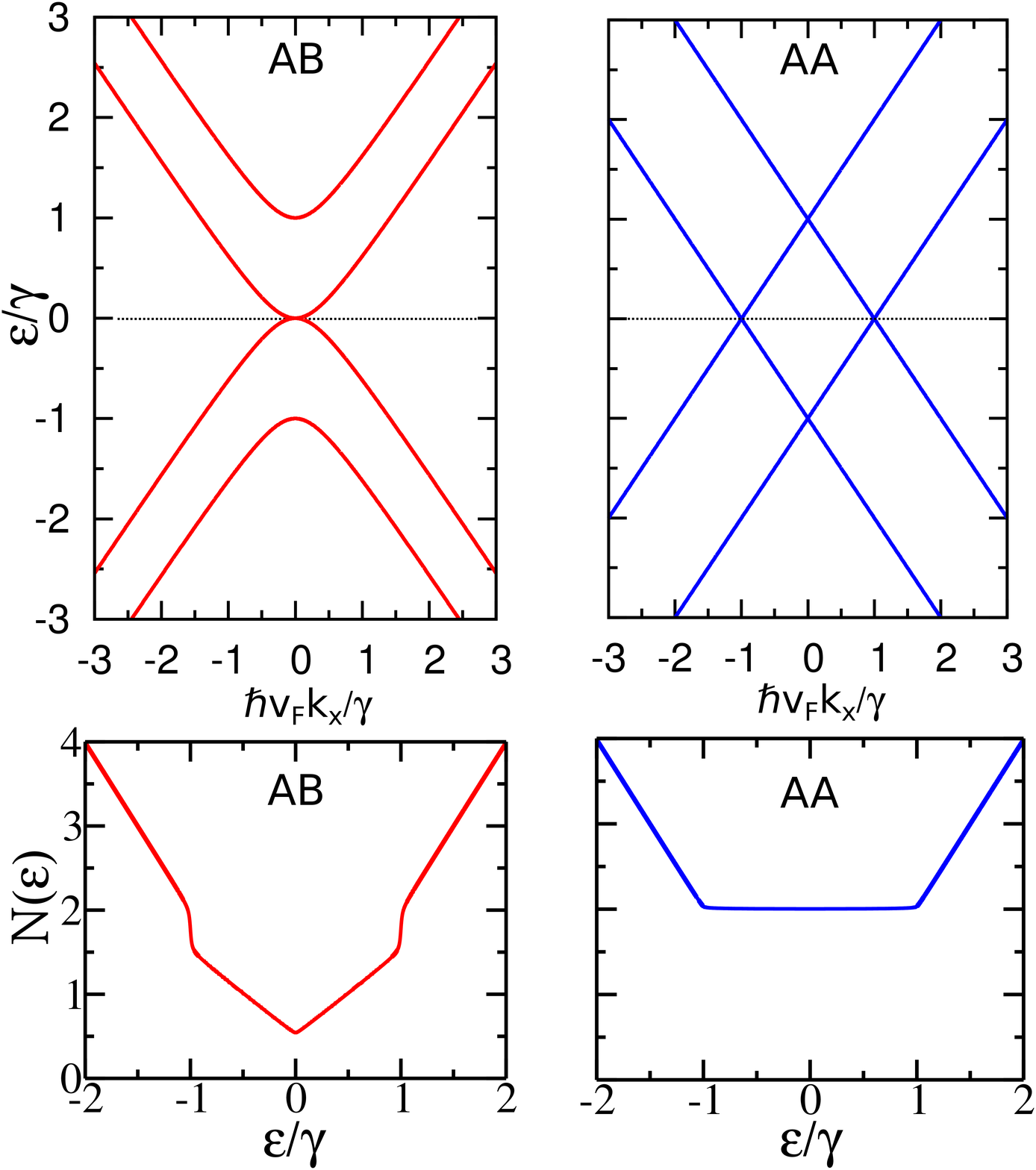}
\caption{(Color online)
Top: Low energy dispersion for a bilayer with Bernal AB-stacking (left) and
 AA-stacking (right). Bottom: Low energy density of states [in units of $2\gamma/\pi(\hbar v_F)^2$] for AB- and AA-stacked bilayers, left and right frames
respectively.
}
\label{fig:AA-ABEnergy}
\end{figure}

For AA stacking, the single spin Hamiltonian is given by
\begin{align}\label{Hamiltonian}
H&=-t\sum_{\bm{n,{\delta}}}\left(b^\dagger_{1\, \bm{n+\delta}} a_{1\, \bm{n}}+H.c.\right)-t\sum_{\bm{n, \delta}}\left(b^\dagger_{2\, \bm{n+\delta}} a_{2\, \bm{n}}+H.c.\right)\notag\\
&+\gamma\sum_{\bm{n}} \left(a^\dagger_{2\, \bm{n}} a_{1\, \bm{n}}+b^\dagger_{2\, \bm{n}}b_{1\, \bm{n}}+H.c.\right).
\end{align}
The first two terms are the nearest-neighbour intralayer hopping terms for electrons to move within a given plane with hopping energy $t\sim 3$ eV.  
The two planes are indexed 1 and 2.  As a consequence of the geometry of the honeycomb lattice, each sheet has two inequivalent atoms labelled A and B.  
The operator $a_{i\,\bm{n}}$ annihilates an electron 
which is on an A-atom site with site label $\bm{n}$ in the graphene sheet indexed by $i$. The label $\bm{n}$ indexes the sites of the
triangular Bravais lattice.
Conversely, $b^\dagger_{i\,\bm{n}+\bm{\delta}}$ creates an electron in sheet $i$ on the neighboring site at the position $\bm{n}+\bm{\delta}$, 
where $\bm{\delta}$ is one of three possible nearest-neighbour vectors given by $\bm{\delta}_1=-(\bm{a}_1+\bm{a}_2)/3$, 
$\bm{\delta}_2=(2\bm{a}_1-\bm{a}_2)/3$ and $\bm{\delta}_3=-(\bm{a}_1-2\bm{a}_2)/3$. The primitive vectors of the triangular sublattice are 
$\bm{a}_1=(a\sqrt{3}/2,a/2)$ and $\bm{a}_2=(a\sqrt{3}/2,-a/2)$, where $|\bm{a}_1|=|\bm{a}_2|=\sqrt{3}a_{cc}$ with $a_{cc}$ the 
shortest carbon-carbon distance.  The third term in Eqn.~\eqref{Hamiltonian} corresponds to the interlayer hopping between graphene sheets.  
The hopping parameter between an A (B) site in one layer and the nearest A (B) site in the other layer is given by $\gamma$ which is reported to be
about 0.2 eV\cite{Xu:2010,Lobato:2011}, which differs in AB-stacked bilayer graphene where it is closer to 0.4 eV.
There is also a possibility to hop between an A (B) site in one layer to a B (A) site in the other layer; however, these hopping 
energies are very small\cite{Charlier:1992,Lobato:2011} and thus ignored in our model.  The Hamiltonian given in Eqn.~\eqref{Hamiltonian} 
transforms to $k$ space in the usual way\cite{Mahan:1990} and can be written in the following matrix representation:
\begin{equation}\label{HamiltonianMatrix}
\hat{H}=
\left(\begin{array}{cccc} 
0 & 0 & \gamma & f(\bm{k}) \\ 
0 & 0 & f^*(\bm{k}) & \gamma \\
\gamma & f(\bm{k}) & 0 & 0 \\
f^*(\bm{k}) & \gamma & 0 & 0
\end{array}\right),
\end{equation}
where $f(\bm{k})=-t\sum_{\bm{\delta}}\,e^{i\bm{k\cdot\delta}}$ and we have used the eigenvector 
$\Psi=(a_{1\,\bm{k}},b_{2\,\bm{k}},a_{2\,\bm{k}},b_{1\,\bm{k}})$ following the notation of McCann\cite{McCann:2006}.  
The band structure is given by the eigenvalues of this matrix.  Reflecting the fact that there are now four atoms per unit cell, 
we obtain the following four energy bands:
\begin{equation}\label{BandStructure}
\varepsilon_\alpha(\bm{k})=\pm[|f(\bm{k})|+(-)^\alpha\gamma ],
\end{equation} 
where $\alpha=1$ and 2 and $|f(\bm{k})|$ is the energy dispersion for a single sheet of graphene.  We essentially have two copies of the band 
structure of monolayer graphene, one shifted by $-\gamma$ and the other by $+\gamma$, or bonding and antibonding bands, 
and indeed we will see that this provides part of the physics
which enters the dynamical conductivity.   As our interest is to understand the conductivity at
low energies, we choose to expand $f(\bm{k})$ around the $K$ point of the Brillouin Zone 
to obtain $f(\bm{k})=\hbar v_F k e^{i\theta}$, where $v_F=\sqrt{3}ta/2\hbar$ and $\theta$ is the $k$-space angle around the $K$ point.  
The low energy band structure can be seen in Fig.~\ref{fig:AA-ABEnergy} where it is compared to that of the familiar Bernal stacking.
As the physics of the conductivity associated with the $K'$ will be the same as for the $K$ point, it is sufficient to work only
about the one $K$ point in what follows and multiply the result by a factor of two for the so-called valley degeneracy associated with the
two $K$ points per unit cell.

We can also provide, as others have shown\cite{Ando:2011}, an analytic expression for the total double spin density of states, $N(\varepsilon)$:
\begin{align}\label{DOSAnalytic}
N(\varepsilon)&=\frac{2\,\gamma}{\pi(\hbar v_F)^2}\bigg[\bigg|\frac{\varepsilon}{\gamma}-1\bigg|+\bigg|\frac{\varepsilon}{\gamma}+1\bigg|\bigg].
\end{align}
which results from the sum of two Dirac cone density of states shifted relative to each other by $2\gamma$.
A plot of the low energy density of states in units of $2\gamma/\pi(\hbar v_F)^2$ for AA-stacked bilayer graphene
is shown in Fig. \ref{fig:AA-ABEnergy} and is contrasted with that for AB-stacking\cite{Nicol:2008}.

With our Hamiltonian, it is straightforward to determine the Green's function $\hat{G}(z)$ from $\hat{G}^{-1}(z)=z\hat{I}-\hat{H}$.  Thus
\begin{equation}\label{GreensInverse}
\hat{G}^{-1}(z)=
\left(\begin{array}{cccc} 
z & 0 & -\gamma & -f(\bm{k}) \\ 
0 & z & -f^*(\bm{k}) & -\gamma \\
-\gamma & -f(\bm{k}) & z & 0 \\
-f^*(\bm{k}) & -\gamma & 0 & z
\end{array}\right).
\end{equation}
The only elements of the Green's function that contribute to our final expressions for longitudinal and perpendicular optical 
conductivity are $G_{11}$, $G_{12}$, $G_{13}$ and $G_{14}$.  We will only show these elements explicitly:
\begin{equation}\label{G_11}
G_{11}(z)=\frac{z^3-z(\gamma^2+|f(\bm{k})|^2)}{(z-\varepsilon_1)(z+\varepsilon_1)(z-\varepsilon_2)(z+\varepsilon_2)},
\end{equation}
\begin{equation}\label{G_12}
G_{12}(z)=\frac{2\gamma zf(\bm{k})}{(z-\varepsilon_1)(z+\varepsilon_1)(z-\varepsilon_2)(z+\varepsilon_2)},
\end{equation}
\begin{equation}\label{G_13}
G_{13}(z)=\frac{z^2\gamma+\gamma|f(\bm{k})|^2-\gamma^3}{(z-\varepsilon_1)(z+\varepsilon_1)(z-\varepsilon_2)(z+\varepsilon_2)},
\end{equation}
and
\begin{equation}\label{G_14}
G_{14}(z)=\frac{z^2f(\bm{k})+\gamma^2f(\bm{k})-f(\bm{k})|f(\bm{k})|^2}{(z-\varepsilon_1)(z+\varepsilon_1)(z-\varepsilon_2)(z+\varepsilon_2)}.
\end{equation}

The finite frequency conductivity is calculated through the standard procedure of using the Kubo formula\cite{Mahan:1990},
where the  conductivity is written in terms of the retarded current-current correlation function. From this the real
part of the conductivity  can be written as\cite{Nicol:2008}
\begin{align}\label{CondIntegral}
\sigma_{\alpha\beta}(\Omega)&=\frac{N_f\,e^2}{2\Omega}\int_{-\infty}^{\infty}\frac{d\omega}{2\pi}\left[f(\omega-\mu)-f(\omega+\Omega-\mu)\right]\notag\\
&\times\int\frac{d^2 k}{(2\pi)^2}\,\text{Tr}\left[\hat{v}_\alpha\hat{A}(\omega+\Omega,\bm{k})\hat{v}_\beta\hat{A}(\omega,\bm{k})\right],
\end{align} 
where we have used the spectral function representation of the Green's function,
 \begin{equation}\label{SpectralDefn}
\hat{G}_{ij}(z)=\int_{-\infty}^\infty \frac{d\omega}{2\pi}\,\frac{\hat{A}_{ij}(\omega)}{z-\omega}.
\end{equation}
Here $\alpha$ and $\beta$ represent the spatial coordinates $x$,$y$,$z$,
$N_f$ is a degeneracy factor, $f(x)=1/[\text{exp}(x/T)+1]$ is the Fermi function for temperature $T$ and $\mu$ is the chemical potential taken to
be positive here but to accommodate for negative values, $\mu$ just needs to replaced by $|\mu|$ everywhere. Note that we will usually take $\hbar=1$ when referring to the relationship between energy and frequency and restore it when necessary. 
For our results, we show only the $T=0$ case.  For the longitudinal in-plane conductivity, $\sigma_{xx}(\Omega)$, $\hat{v}_\alpha=\hat{v}_\beta=\hat{v}_x$ where
\begin{equation}
\hat{v}_x=
\left(\begin{array}{cccc} 
0 & 0 & 0 & v_F \\ 
0 & 0 & v_F & 0 \\
0 & v_F & 0 & 0 \\
v_F & 0 & 0 & 0
\end{array}\right).
\end{equation} 
The velocity operator can be evaluated from a Peierls substitution as demonstrated in Ref.~\cite{Nicol:2008} or from
$\hbar\hat v_{x}=\partial \hat H/\partial k_x$.
We can then evaluate the trace, drop the terms that will vanish upon averaging over angle and obtain an expression dependent on the two 
spectral functions $A_{11}$ and $A_{13}$.  In the zero temperature limit, the real part of the longitudinal conductivity 
$\sigma_{xx}(\Omega)$ is then
\begin{align}\label{ConductivitySpectral}
\sigma_{xx}(\Omega)&=\frac{N_f\,e^2}{2\Omega}\int_{\mu-\Omega}^\mu\frac{d\omega}{2\pi}
\int\frac{d^2 k}{(2\pi)^2}\,4v_F^2\notag\\
&\times\bigg[A_{11}(\omega+\Omega)A_{11}(\omega)+A_{13}(\omega+\Omega)A_{13}(\omega)\bigg].
\end{align}
In keeping with the low energy expansion of $f(\bm{k})$ about a single $K$ point, the 
integral over $k$, which in general is over the first Brillouin Zone, is now taken as an integral over a single $K$ point.  
The degeneracy factor is thus $N_f=g_s\,g_v$, where $g_s=2$ to account for the sum over spin which has been ignored up until now and $g_v=2$ 
to account for a sum over the $K$ and $K'$ points of the Brillouin Zone.  Furthermore, the upper limit of the 
$k$ integral is taken to be a large cutoff value typical of momentum associated with the large bandwidth.    
It is convenient to scale Eqn.~\eqref{ConductivitySpectral} by the constant background conductivity of a single sheet 
of graphene\cite{Gusynin:2006} given by $\sigma_0=e^2/4\hbar$.  All that remains before we can calculate our conductivity is to 
specify the necessary spectral function elements.  Given our expressions for $G_{11}$ and $G_{13}$ and Eqn.~\eqref{SpectralDefn}, we obtain
\begin{equation}\label{A_11}
A_{11}=\frac{\pi}{2}\left[\delta(\omega-\varepsilon_1)+\delta(\omega+\varepsilon_1)+
\delta(\omega-\varepsilon_2)+\delta(\omega+\varepsilon_2)\right]
\end{equation}
and
\begin{equation}\label{A_13}
A_{13}=\frac{\pi}{2}\left[\delta(\omega+\varepsilon_1)-\delta(\omega-\varepsilon_1)+\delta(\omega-\varepsilon_2)-\delta(\omega+\varepsilon_2)\right].
\end{equation} 
For our numerical work, we use the Lorentzian representation of the delta function, $\delta(x)=(\eta/\pi)/[\eta^2+x^2]$, with a 
broadening of $\eta=0.01\gamma$. The broadening is manifest in the optical conductivity as an effective transport scattering rate of $1/\tau_{imp}=2\eta$ 
due to the convolution of the two Lorentzian functions in the conductivity formula.
 
We can also examine the perpendicular conductivity, $\sigma_{zz}(\Omega)$, associated with transport perpendicular to the graphene sheets.  
In Eqn.~\eqref{CondIntegral}, our velocity operator is now ($\hat{v}_\alpha=\hat{v}_\beta=\hat{v}_z$)
\begin{equation}\label{Velocity matrix}
\hat{v}_z=
\left(\begin{array}{cccc} 
0 & 0 & v^* & 0 \\ 
0 & 0 & 0 & v \\
v & 0 & 0 & 0 \\
0 & v^* & 0 & 0
\end{array}\right),
\end{equation}
where $v=i\,\gamma\,d/\hbar$ with $d$ the interlayer distance. $d$ is about 3.6~\AA\, and 3.3~\AA\, for AA- and AB-stacking
respectively\cite{Xu:2010,Lobato:2011}.  
 This leads to the real part of the zero temperature perpendicular conductivity:
\begin{align}\label{Cond-Perp-def}
\sigma_{zz}(\Omega)&=\frac{N_f\,e^2}{2\Omega}\int_{\mu-\Omega}^\mu\frac{d\omega}{2\pi}
\int\frac{d^2 k}{(2\pi)^2}\,4\,|v|^2\notag\\
&\times\bigg[\,A_{11}(\omega+\Omega)A_{11}(\omega)-\,A_{12}^*(\omega+\Omega)A_{12}(\omega)\notag\\
&-\,A_{13}(\omega+\Omega)A_{13}(\omega)+\,A_{14}^*(\omega+\Omega)A_{14}(\omega)\bigg].
\end{align}
with $A_{11}$ and $A_{13}$ given by Eqns.~\eqref{A_11} and \eqref{A_13}, respectively, and
\begin{align}\label{A_12}
A_{12}=\frac{\pi\,f(\bm{k})}{2\varepsilon}[-\delta(\omega-\varepsilon_1)-\delta(\omega+\varepsilon_1)&+\delta(\omega-\varepsilon_2)\notag\\
&+\delta(\omega+\varepsilon_2)]
\end{align}
and
\begin{align}\label{A_14}
A_{14}=\frac{\pi\,f(\bm{k})}{2\varepsilon}[\delta(\omega-\varepsilon_1)-\delta(\omega+\varepsilon_1)&+\delta(\omega-\varepsilon_2)\notag\\
&-\delta(\omega+\varepsilon_2)],
\end{align}
with $\varepsilon=|f(\bm{k})|$.

\section{Results for AA-stacked bilayer}
 
\begin{figure}
\includegraphics[width=0.8\linewidth]{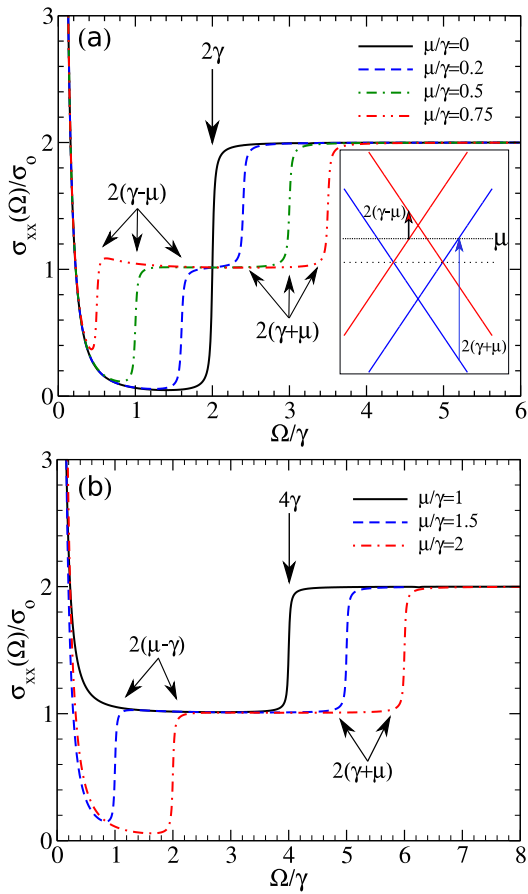}
\caption{(Color online)
(a) Real part of the longitudinal frequency-dependent optical conductivity $\sigma_{xx}(\Omega)$ of AA-stacked bilayer graphene normalized to 
that of a single graphene sheet for the case of $\mu<\gamma$.  Onset of absorption at $2|\gamma-\mu|$ and $2(\gamma+\mu)$ 
correspond the availability of allowed transitions within the AA band structure shown in the inset.
Inset: Low energy band structure around the $K$ point for the case of $\mu<\gamma$.
Transitions can only occur within the same colored bands. For the blue bands, the lowest transition between the upper and lower 
cones occurs at $\Omega=2(\gamma+\mu)$, shown by the blue arrow, and for the red bands, 
the lowest transition between cones occurs at $\Omega=2(\gamma-\mu)$, shown by the red arrow.  Below these frequencies, 
only intraband transitions within the same cone for states about the chemical potential occur.  (b) The case of $\mu>\gamma$.  
}
\label{fig:Conductivity-AA}
\end{figure}

Here we present results for the longitudinal conductivity which is obtained by
evaluating Eqn.~\eqref{ConductivitySpectral} numerically using the Lorentzian form in place of the delta function and taking the broadening
parameter $\eta=0.01\gamma$. Fig.~\ref{fig:Conductivity-AA} shows curves for the case of $\mu<\gamma$ 
and $\mu>\gamma$ (top and bottom frames, respectively).  For the case of charge neutrality ($\mu=0$), the conductivity displays a Drude response
at low frequency due to intraband transitions and a flat interband absorption which commences at $2\gamma$. This is quite unlike the
case of monolayer graphene which would have had a flat interband response at all frequencies. Here the response is not simply that of
two monolayers as one might naively think. This is because the AA-stacked bands are essentially two decoupled graphene monolayer bands which represent bonding and antibonding bands and are
shifted relative to each other\cite{Ando:2011} (as emphasized in the inset where the bands are identified with
different colors). The matrix elements for the longitudinal 
conductivity only allow for transitions between like-colored bands.\cite{Xu:2010}  The transitions must be vertical as the photon is a momentum $q\sim 0$ probe. At charge neutrality, the
minimum interband transition that is not Pauli blocked is then $2\gamma$ and intraband transitions can also now occur due to the Fermi level being
located away from the Dirac point of the monolayer band, this latter feature is not present in previous work\cite{Xu:2010}.  
This emphasizes that  knowledge of the band structure as shown in Fig.~\ref{fig:AA-ABEnergy}
is insufficient (or may be misleading) to the determination of the allowed absorption transitions and that the matrix elements which know
about effects of chirality and bonding/antibonding are important as well.  

For finite chemical potential, the Drude persists but now there are two pieces to the
interband absorption which onset at $2|\gamma-\mu|$ and $2(\gamma+\mu)$.  The behavior is different for $\mu<\gamma$ versus $\mu>\gamma$.
For the case of $\mu<\gamma$, the Drude conductivity remains completely unchanged and its weight is set by the value of $\gamma$.
The interband edge that was at $2\gamma$ in the $\mu=0$ case is now split into two edges moving to lower and higher frequency
and associated with the onset of allowed transitions for a single monolayer band structure shifted up or down by $\gamma$, respectively.
This is emphasized by the AA-stacked band structure shown in the inset 
of Fig.~\ref{fig:Conductivity-AA}(a) where transitions can only occur within the same colored bands.
For the monolayer dispersion shifted down by $\gamma$ (blue bands) the lowest interband transition from an occupied state to
an unoccupied state occurs at $\Omega=2(\gamma+\mu)$.  For the monolayer dispersion shifted up by $\gamma$ (red bands), 
the lowest transition between cones occurs at $\Omega=2|\gamma-\mu|$.  Thus at low enough energy, the finite frequency conductivity
displays the universal background absorption of a monolayer $\sigma_0$ but at higher frequency there is a step-up to a flat universal background
at $2\sigma_0$ and the transition between these steps is tunable with the charge doping.  

For the case of $\mu=\gamma$
as shown in Fig.~\ref{fig:Conductivity-AA}(b), 
the lower edge has disappeared and there is a background conductivity of $\sigma_0$ for $\Omega<4\gamma$
and $2\sigma_0$ for $\Omega>4\gamma$. However, in this case, the Drude component still remains at very low frequency. For $\mu>\gamma$ as shown in 
Fig.~\ref{fig:Conductivity-AA}(b), the lower edge reappears, showing the double step in universal conductivity value and now both edges move to higher
frequency with increased $\mu$. As the area of the conductivity is conserved, the lost weight at finite frequency is found in the Drude which
now increases with $\mu$ as one finds in the monolayer case. These characteristic features of the AA-stacked bilayer are quite different from
the case of Bernal stacking\cite{Nicol:2008} and are not at all the expectation of twice the monolayer conductivity either. The presence of the Drude at charge neutrality
in the AA-stacked case is different from the monolayer and AB-stacked bilayer where no such feature exists for $\mu=0$. These special
features of AA-stacked graphene might prove useful for applications where optical response is tuned by doping (or gating) to be like a switch
with three settings: off or 0, on at half setting ($\sigma_0$) and on at full setting ($2\sigma_0$). Use of tuning by gating has been
demonstrated for graphene terahertz modulators where the intraband transitions are used in this case.\cite{Sensale:2012}
Another significance of the result is that for the spectral range between $2|\mu-\gamma|$ and $2(\gamma+\mu)$, one might not be able to 
separate monolayer from AA-stacked bilayers by optics alone. Overall, the dynamical conductivity is quite distinct from that of AB-stacked bilayer
where no such steps occur and the conductivity is not flat in the low frequency spectral range.\cite{Abergel:2007,Nicol:2008}

These results are embodied by a closed algebraic formula for the real part of the longitudinal conductivity for AA bilayer which can be derived
at zero temperature from Eqn.~\eqref{ConductivitySpectral}:  
 \begin{align}\label{ConductivityAnalytic}
\frac{\sigma_{xx}(\Omega)}{\sigma_0}=8\,\delta(\Omega)\,\text{max}(\mu,\gamma)&+\Theta[\Omega-2|\mu-\gamma|]\notag\\
&+\Theta[\Omega-2(\mu+\gamma)].
\end{align}
If the delta function here is replaced by a Lorentzian with broadening of $2\eta$, then this formula gives an excellent representation of the
numerical results in Fig.~\ref{fig:Conductivity-AA}.
It is also possible to derive an expression for the imaginary part of the longitudinal conductivity which is Kramers-Kronig-related to 
Eqn.~\eqref{ConductivityAnalytic} by the relation
\begin{equation}\label{KramersKronig}
\sigma^{\prime\prime}(\Omega)=-\frac{2\Omega}{\pi}\,\mathcal{P}\int_0^\infty \frac{\sigma^\prime(\omega)}{\omega^2-\Omega^2} d\omega
\end{equation}
with $\sigma^\prime(\omega)$, the real part of the conductivity given by Eqn.~\eqref{ConductivityAnalytic}. Hence,
\begin{align}\label{ImagCond}
&\frac{\sigma^{\prime\prime}_{xx}(\Omega)}{\sigma_0}=\frac{8}{\pi\Omega}\text{max}(\gamma,\mu)\notag\\
&+\frac{1}{\pi}\bigg[\text{ln}\bigg|{\Omega-2|\gamma-\mu|\over \Omega+2|\gamma-\mu|}\bigg|
+\text{ln}\bigg|{\Omega-2(\gamma+\mu)\over \Omega+2(\gamma+\mu)}\bigg|\bigg].
\end{align}
If $\delta(\Omega)$ in the real part is replaced by a Lorentzian
form as we have discussed, then instead of the Kramers-Kronig 
transformation of $\delta(\Omega)$ to $1/(\pi\Omega)$, we have
$\Gamma/[\pi(\Omega^2+\Gamma^2)]$ transforming to $\Omega/[\pi(\Omega^2+\Gamma^2)]$ where $\Gamma=2\eta$ represents the transport scattering rate rather than
the quasiparticle scattering rate $\eta$ that enters the broadened spectral
functions $A(\bm{k},\omega)$. 

\begin{figure}
\includegraphics[width=0.8\linewidth]{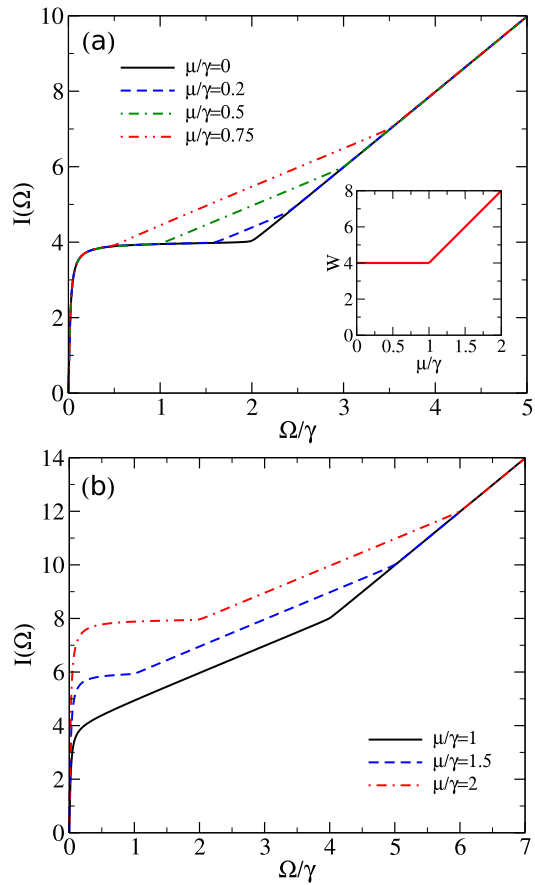}
\caption{(Color online)
Partial optical sum $I(\Omega)$ in units of $\gamma$ for various values of chemical potential with (a) $\mu<\gamma$
and (b) $\mu>\gamma$.
Inset: The evolution of the positive frequency spectral weight W, found under half the delta function in 
the analytic solution for the conductivity, Eqn.~\eqref{ConductivityAnalytic} as a function of chemical potential. 
}
\label{fig:PartialOpSum}
\end{figure}

The issue of optical spectral weight redistribution with variation in chemical potential can be addressed globally by introducing the partial optical sum:
\begin{equation}\label{PartialOpSum}
I(\Omega)=\int_{0^+}^\Omega \frac{\sigma(\omega)}{\sigma_0}d\omega,
\end{equation}
which is defined as the area under the conductivity graph up to energy $\Omega$. For the real part of the 
longitudinal conductivity, $I(\Omega)$ is shown for various values of $\mu/\gamma$ in Fig.~\ref{fig:PartialOpSum}.  
In all cases, at sufficiently high frequency the integrated spectral weight returns to the $\mu=0$ value 
(solid black curve of Fig.~\ref{fig:PartialOpSum}(a)).  The inset of Fig.~\ref{fig:PartialOpSum} shows the spectral weight of the 
delta function in the analytic solution, Eqn.~\ref{ConductivityAnalytic}, where we have taken only half the weight of the delta function 
as only half the function is present for positive frequency.  The functions converge to the $\mu=0$ case 
at frequencies above $\Omega=2(\gamma+\mu)$ once most of the spectral weight from the Drude contribution is 
integrated as well as the contribution from the interband edges. These curves once again provide an interesting differentiation
between the $\mu<\gamma$ regime, where the Drude weight remains the same and the lower energy kink moves to lower $\Omega$,
and the $\mu>\gamma$ case where the low frequency part of the partial sum increases with $\mu$ and the low energy kink moves
to higher $\Omega$. In principle, such a quantity could allow for an experimental determination of $\gamma$ based on the
transition from the behavior of one regime to the other.

\begin{figure}
\includegraphics[width=0.8\linewidth]{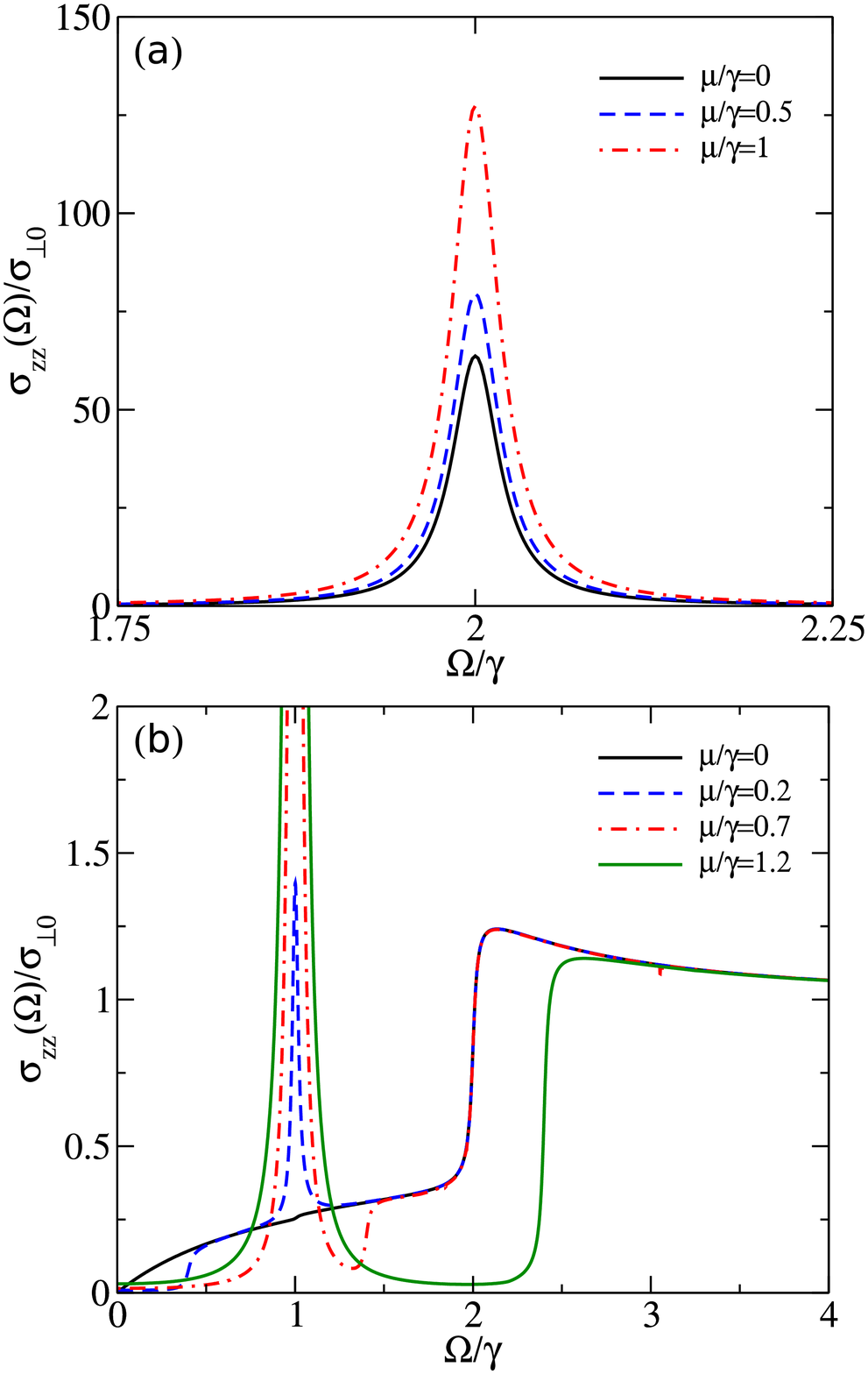}
\caption{(Color online)
(a) Real part of the perpendicular conductivity for AA-stacked bilayer graphene for various values of $\mu$.  The perpendicular conductivity 
is negligible everywhere except near $\Omega=2\,\gamma$ where there is a sharp peak (note the axes of this plot). Here,
the transport scattering rate is $\Gamma=2\eta$, where $\eta=0.01\gamma$
in all of our numerical work. 
(b) Real part of the perpendicular conductivity for AB-stacked bilayer graphene.  A strong peak occurs $\gamma$ for finite $\mu$
and an absorption edge occurs at 2 max($\mu,\gamma$).
}
\label{fig:Conductivity-Perp}
\end{figure}

Turning to the perpendicular conductivity, we show the response for the AA- and the AB-stacked bilayer in Fig.~\ref{fig:Conductivity-Perp}.
The AA-stacked graphene has a strong absorption associated with $2\gamma$ which is finite at charge neutrality and increases with doping.
It is also possible to derive a closed form algebraic formula for the perpendicular conductivity.  For AA-stacking we obtain:
\begin{align}\label{Cond-Perp-AA}
\frac{\sigma_{zz}(\Omega)}{\sigma_{\perp 0}}=\frac{4}{\gamma}\delta(\Omega-2\,\gamma)\bigg[(\gamma-\mu)^2\Theta(\gamma-\mu)+2\gamma\mu\bigg]
\end{align}
and by Kramers-Kronig transformation
\begin{align}\label{Cond-Perp-AA-Imag}
\frac{\sigma_{zz}^{\prime\prime}(\Omega)}{\sigma_{\perp 0}}=\frac{8\Omega}{\gamma\pi(\Omega^2-4\gamma^2)}\left[(\gamma-\mu)^2\Theta(\gamma-\mu)+2\gamma\mu\right],
\end{align}
for the real and imaginary parts, respectively,
where $\sigma_{\perp 0}=(e^2/4\hbar)\cdot(\gamma\,d/\hbar\,v_F)^2$. The physics of this case is as follows. For in-plane
conductivity, charge carriers must hop from one sublattice to the other to produce a current, thus in the absorption
process, interband transitions are between two bands, each of which reflects the two sublattices by having different
chirality label. Hence the transitions shown in the inset of Fig.~\ref{fig:Conductivity-AA}(a) are between two cones
which have opposite chirality but the same bonding or antibonding wavefunctions.
 However, for the interlayer current in the AA-stacked case, the carriers  hop between
the A(B)-sublattice of one plane to the A(B)-sublattice of the other plane and as a result absorptive transitions for
this form of transport will only occur between bands of the same chirality
but different bonding which in reference to 
the inset of Fig.~\ref{fig:Conductivity-AA}(a) would be vertical arrows connecting the parallel bands in this case. As
these arrows are always of length $2\gamma$, there is only one very strong absorption peak at $2\gamma$.

Similar analytical results for AB-stacking at finite doping are, to our knowledge, not in the literature
and so we will provide it here for comparison 
(the equivalent
form for AB-stacked in-plane conductivity has been given previously in the literature\cite{Nicol:2008,Koshino:2009}).  
Some numerical work along with some analytical analysis has been done previously and our results are in
agreement with those works.\cite{Nilsson:2008,Ando:2009} In particular, Ando and Koshino considered polarization effects\cite{Ando:2009}
which we do not include here. The AB-stacked perpendicular response  seen in Fig. \ref{fig:Conductivity-Perp}(b) is quite different from the AA case. Absorption
occurs at all frequencies but is on the scale of $\sigma_{\perp 0}$. An absorption edge occurs at $2\gamma$
similar to the AA case, although it is weaker by comparison and continues on to higher frequency
as it results from transitions between the lowest and highest energy bands of the AB case in Fig.~\ref{fig:AA-ABEnergy}, 
which are the
bonding and antibonding bands of the A-B dimer strongly coupled by hopping $\gamma$.\cite{McCann:2012} 
Moreover, absorption is seen at all frequencies and at finite doping a very strong peak
occurs at $\gamma$ much as is seen in the in-plane conductivity for the AB-stacked bilayer\cite{Nicol:2008}.
It is
not entirely surprising that the perpendicular conductivity displays elements of the in-plane conductivity
with the same physical origin\cite{Nicol:2008}. The lower energy bands represent
hopping between the non-dimer A and B sites in the two planes which must occur by first hopping
in the plane to the neighbour site which is part of a dimer, hopping up the dimer bond and then over to
the non-dimered site in the second plane.\cite{McCann:2012} The AA-stacked case does not have this element.
For AB-stacking, we derive an analytic formula for the perpendicular conductivity which is:
\begin{align}\label{Cond-Perp-AB}
\frac{\sigma_{zz}(\Omega)}{\sigma_{\perp 0}}=\bigg[\frac{\Omega}{2(\Omega+\gamma)
}+\frac{\Omega}{2(\Omega-\gamma)}&\Theta(\Omega-2\gamma)\bigg]\Theta(\Omega-2\mu)\notag\\
&+c(\mu)\delta(\Omega-\gamma),
\end{align}
where
\begin{align}\label{c-mu}
c(\mu)=\frac{\mu(\gamma+\mu)}{\gamma}&-\frac{\gamma}{2}\text{ln}\frac{2\mu+\gamma}{\gamma}\notag\\
&+\bigg[\frac{\mu(\gamma-\mu)}{\gamma}+\frac{\gamma}{2}\text{ln}\frac{2\mu-\gamma}{\gamma}\bigg]\Theta(\mu-\gamma).
\end{align}
This formula also agrees quite well with the numerical work shown in
Fig.~\ref{fig:Conductivity-Perp}(b) for various values of the chemical potential, 
provided the delta function is rewritten as a Lorentzian
with broadening of $\Gamma=2\eta$.
The expression for the imaginary part is given as:
\begin{align}\label{Cond-Perp-AB-Imag}
\frac{\sigma_{zz}^{\prime\prime}(\Omega)}{\sigma_{\perp 0}}=\frac{2\Omega}{\pi(\Omega^2-\gamma^2)}c(\mu)+&\frac{\Omega}{2\pi}\bigg[\frac{1}{\Omega-\gamma}\text{ln}\bigg|\frac{2\text{max}(\gamma,\mu)-\Omega}{2\mu+\Omega}\bigg|\notag\\
&-\frac{1}{\Omega+\gamma}\text{ln}\bigg|\frac{2\text{max}(\gamma,\mu)+\Omega}{2\mu-\Omega}\bigg|\notag\\
&+\frac{2\gamma}{\Omega^2-\gamma^2}\text{ln}\bigg|\frac{2\mu+\gamma}{2\text{max}(\mu,\gamma)-\gamma}\bigg|\bigg],
\end{align}
where $c(\mu)$ is given by Eqn.~\eqref{c-mu}.

We can also examine the effect of adding a bias between the two layers.  To do this, we need to include an additional term in our Hamiltonian given by Eqn.~\eqref{Hamiltonian} of the form\cite{Nicol:2008}
\begin{align}\label{Ham-Bias}
H^\prime =&\frac{1}{2}\Delta\sum_{\bm{n}}(a^\dagger_{1\bm{n}}a_{1\bm{n}}+b^\dagger_{1\bm{n}+\bm{\delta}_1}b_{1\bm{n}+\bm{\delta}_1})\notag\\
&-\frac{1}{2}\Delta\sum_{\bm{n}}(a^\dagger_{2\bm{n}}a_{2\bm{n}}
+b^\dagger_{2\bm{n}+\bm{\delta}_1^\prime}b_{2\bm{n}+\bm{\delta}_1^\prime}).
\end{align}
This bias raises the energy on the lower plane by $+\Delta/2$ and lowers the energy on the upper plane by $-\Delta/2$ providing an overall bias of $\Delta$. 
For the case of AB-stacking, this introduces a gap in the energy dispersion\cite{McCann:2006} and provides interesting features to the conductivity\cite{Nicol:2008}.  
For AA-stacked bilayer, this gives the energy dispersion $\varepsilon_{\alpha}(\bm{k})=\pm[\varepsilon+(-)^\alpha\sqrt{\gamma^2+\Delta^2/4}]$,
 where $\alpha=\pm 1$.  We can see that this is equivalent to a renormalization of the interlayer hopping parameter of the unbiased system to a value 
$\gamma^\prime$ such that $\gamma^\prime=\sqrt{\gamma^2+\Delta^2/4}$ and will therefore introduce no new features into the conductivity.

\section{Conductivity of an AAA-stacked trilayer}

\begin{figure}
\includegraphics[width=0.8\linewidth]{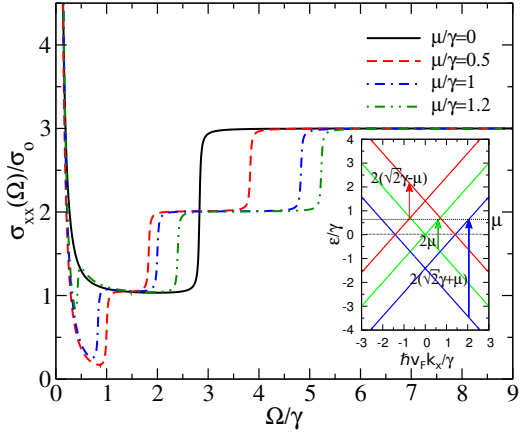}
\caption{(Color online)
Conductivity for AAA-stacked trilayer graphene. Interband absorption edges appear at 
$\Omega=2\mu,\,2|\sqrt{2}\gamma-\mu|\,\text{and}\,2(\sqrt{2}\gamma+\mu)$. These transitions are shown on the band structure given in the
inset.
}
\label{fig:Conductivity-AAA}
\end{figure}

These ideas can also be extended to trilayer graphene.  For the case of AAA-stacked trilayer graphene, 
where A (B) sites in each layer are stacked directly in line with the corresponding sites in the other layers, our Hamiltonian now becomes
\begin{align}\label{Ham-AAA}
H&=-t\sum_{i,\bm{n,{\delta}}}\left(b^\dagger_{i\, \bm{n+\delta}} a_{i\, \bm{n}}+H.c.\right)\notag\\
&+\gamma\sum_{\bm{n}} \left(a^\dagger_{2\, \bm{n}} a_{1\, \bm{n}}+b^\dagger_{2\, \bm{n}}b_{1\, \bm{n}}+a^\dagger_{3\, \bm{n}} a_{2\, \bm{n}}+b^\dagger_{3\, \bm{n}}b_{2\, \bm{n}}+H.c.\right),
\end{align}
where $i=1$, 2, 3 indexes each of the three layers. Here, we allow the usual nearest-neighbour intralayer hopping $t$ and 
interlayer hopping $\gamma$ between the neighbouring planes; again, we have ignored hopping from an A(B) site in one 
layer to a B(A) site in another layer as well as hopping from an A1(B1) site to an A3(B3) site as 
these hopping energies are very small\cite{Charlier:1992}. Transforming to $k$ space, we obtain the following matrix representation:
\begin{equation}\label{HamiltonianMatrixAAA}
\hat{H}=
\left(\begin{array}{cccccc} 
0 & 0 & 0 & 0 & \gamma & f(\bm{k}) \\ 
0 & 0 & 0 & \gamma & f^*(\bm{k}) & \gamma \\
0 & 0 & 0 & f(\bm{k}) & \gamma & 0\\
0 & \gamma & f^*(\bm{k}) & 0 & 0 & 0\\
\gamma & f(\bm{k}) & \gamma & 0 & 0 & 0\\
f^*(\bm{k}) & \gamma & 0 & 0 & 0 & 0
\end{array}\right),
\end{equation}
where we have used the eigenvector $\Psi=(a_{1\,\bm{k}},b_{2\,\bm{k}},a_{3\,\bm{k}},b_{3\,\bm{k}},a_{2\,\bm{k}},b_{1\,\bm{k}})$.  
Reflecting the fact that we now have six atoms per unit cell, we obtain the following six energy bands:
\begin{equation}\label{Energy-AAA}
\varepsilon(\bm{k})=\pm|f(\bm{k})|,
\pm[|f(\bm{k})|-\sqrt{2}\gamma], \pm[|f(\bm{k})|+\sqrt{2}\gamma],
\end{equation}
where $|f(\bm{k})|$ is the energy dispersion of monolayer graphene, equal to $\varepsilon=\hbar v_F k$ at low energy. The first two bands, $\pm\varepsilon$, 
are the original graphene bands, the second two bands, indexed $\pm\varepsilon_1(\bm{k})$, and final two bands, 
indexed $\pm\varepsilon_2(\bm{k})$, are monolayer bands shifted by $\mp\sqrt{2}\gamma$, respectively. 
A plot of the band structure can be seen in the inset of Fig.~\ref{fig:Conductivity-AAA}. From this, we see that the trilayer is like the sum of a monolayer
and a bilayer with an interlayer
hopping of $\sqrt{2}\gamma$.

Using the same formalism as before, we can derive an expression for the real part of the zero temperature 
longitudinal conductivity in terms of the spectral functions; we obtain:
\begin{align}\label{CondAAA}
\sigma_{xx}(\Omega)&=\frac{N_f\,e^2}{2\Omega}\int_{\mu-\Omega}^{\mu} \frac{d\omega}{2\pi}\int \frac{d^2k}{(2\pi)^2}\,2 v_F^2\notag\\
&\times[2A_{11}(\omega+\Omega)A_{11}(\omega)+2A_{13}(\omega+\Omega)A_{13}(\omega)\notag\\
&\,\,+4A_{15}(\omega+\Omega)A_{15}(\omega)+A_{22}(\omega+\Omega)A_{22}(\omega)],
\end{align}
where
\begin{align}\label{A_11-AAA}
A_{11}=\frac{\pi}{4}\bigg[2\delta(\omega-&\varepsilon)+2\delta(\omega+\varepsilon)+\delta(\omega-\varepsilon_1)\notag\\
&+\delta(\omega+\varepsilon_1)+\delta(\omega-\varepsilon_2)+\delta(\omega+\varepsilon_2)\bigg],
\end{align}
\begin{align}\label{A_13-AAA}
A_{13}=\frac{\pi}{4}\bigg[-2\delta(\omega-&\varepsilon)-2\delta(\omega+\varepsilon)+\delta(\omega-\varepsilon_1)\notag\\
&+\delta(\omega+\varepsilon_1)+\delta(\omega-\varepsilon_2)+\delta(\omega+\varepsilon_2)\bigg],
\end{align}
\begin{align}\label{A_15-AAA}
A_{15}=\frac{\sqrt{2}\pi}{4}\bigg[-\delta(\omega-\varepsilon_1)+\delta(\omega+\varepsilon_1)+\delta(\omega&-\varepsilon_2)\notag\\
&-\delta(\omega+\varepsilon_2)\bigg],
\end{align}
and
\begin{align}\label{A_22-AAA}
A_{22}=\frac{\pi}{2}\bigg[\delta(\omega-\varepsilon_1)+\delta(\omega+\varepsilon_1)+\delta(\omega&-\varepsilon_2)\notag\\
&+\delta(\omega+\varepsilon_2)\bigg].
\end{align}

Several numerical curves for the conductivity can be seen in 
Fig.~\ref{fig:Conductivity-AAA}.  For finite $\mu$, there are three steps in the conductivity each of value $\sigma_0$, leading
to a 
constant background at high frequency equal to three times that of a single sheet, reflecting the trilayer nature of the system.  
In each case, there is always an absorption edges at $\Omega=2\mu$. The other two edges occur at 
$2|\mu-\sqrt{2}\gamma|$ and $2(\mu+\sqrt{2}\gamma)$, where the former decreases with increasing $\mu$ for $\mu<\sqrt{2}\gamma$ and then
increases for $\mu>\sqrt{2}\gamma$, while the latter always increases with $\mu$. This is a similar pattern to the AA-stacked case
and so the combination of the above confirms that the trilayer acts as the sum of a monolayer plus bilayer. Even at $\mu=0$, the
flat background of $\sigma_0$, due to monolayer behavior, is added to the unusual Drude plus $2\sigma_0$ interband behavior of the $\mu=0$
bilayer shown in Fig.~\ref{fig:Conductivity-AA}, but for effective interlayer hopping of $\sqrt{2}\gamma$. The ability to tune the
flat background in the IR spectral region from $0\to 3\sigma_0$ in steps of $\sigma_0$ by changing the doping is an interesting
feature that could possibly be of some advantage to technological applications. 

Given Eqn.~\eqref{CondAAA},  an analytical expression for $\sigma_{xx}$ of AAA-stacked trilayer graphene may be written down. It has the 
expected form:
\begin{align}\label{CondAAA-Analytic}
\frac{\sigma_{xx}(\Omega)}{\sigma_0}=&4\delta(\Omega)[2\,{\rm max}(\mu,\sqrt{2}\gamma)+\mu]+\Theta[\Omega-2\mu]\notag\\
&+\Theta[\Omega-2(\mu+\sqrt{2}\gamma)]+\Theta[\Omega-2|\mu-\sqrt{2}\gamma|],
\end{align}
which stresses the existence of three decoupled and shifted 
monolayer dispersions where interband transitions are only permitted 
between the corresponding cones shown in the inset of Fig.~\ref{fig:Conductivity-AAA}.

The imaginary part is again found by applying Eqn.~\eqref{KramersKronig} to Eqn.~\eqref{CondAAA-Analytic}.  It is
\begin{align}
&\frac{\sigma_{xx}^{\prime\prime}(\Omega)}{\sigma_0}=\frac{4}{\pi\Omega}[2\text{max}(\mu,\sqrt{2}\gamma)+\mu]+\frac{1}{\pi}\text{ln}\bigg|\frac{\Omega-2\mu}{\Omega+2\mu}\bigg|\notag\\
&+\frac{1}{\pi}\text{ln}\bigg|\frac{\Omega-2(\mu+\sqrt{2}\gamma)}{\Omega+2(\mu+\sqrt{2}\gamma)}\bigg|+\frac{1}{\pi}\text{ln}\bigg|\frac{\Omega-2(\mu-\sqrt{2}\gamma)}{\Omega+2(\mu-\sqrt{2}\gamma)}\bigg|.
\end{align}

\section{AA-stacking with Spin Orbit Coupling}

Finally, to cover a variety of possible scenarios, we examine the effect of spin orbit coupling (SOC) on AA-stacked bilayer graphene.  
Such effects in AA- and AB-stacked bilayers were considered by Prada \emph{et al.}\cite{Prada:2011} in the context of studying systems
which may manifest a topological insulating phase. These authors have studied both the case of SOC in each plane
and SOC in only one  plane, the latter case taken to be a toy model for spin-orbit proximity effect. For our purpose here,
these considerations illustrate the generic features of opening energy gaps in the AA-stacked band structure at $k=0$ and at
the charge neutrality point. Recall from our previous discussion that biasing the bilayer does not open a gap in the AA-stacked case
as it does in the AB-stacked case, but SOC will do so.

\begin{figure}[h!]
\includegraphics[width=0.8\linewidth]{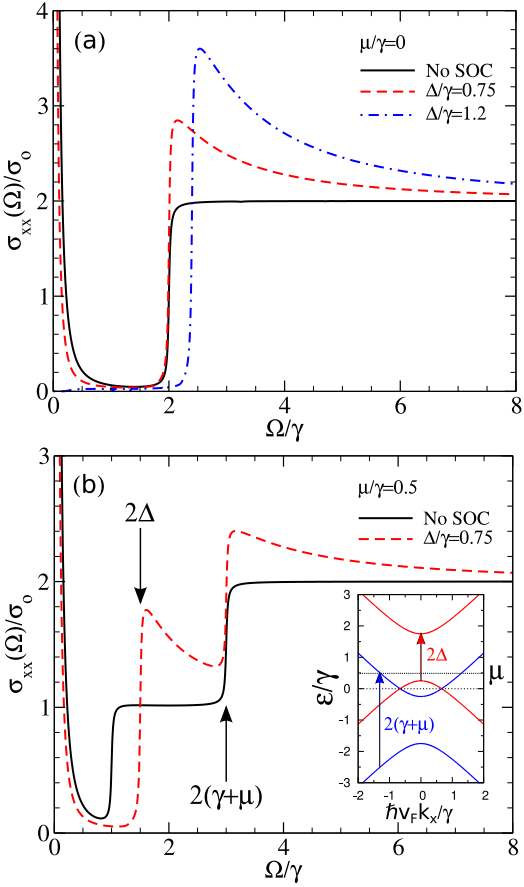}
\caption{(Color online)
Conductivity $\sigma_{xx}(\Omega)$ versus $\Omega/\gamma$ for the case with SOC in both layers. (a) $\mu=0$
and $\Delta/\gamma=0$, 0.75 and 1.2 in each layer. (b) The case of finite doping with $\mu=0.5\gamma$
and $\Delta=0.75\gamma$ with that for $\Delta=0$ shown for comparison. The inset shows the band structure
for this case and the transitions that occur to give the absorption edges seen in the main frame.
}
\label{fig:Conductivity-SOC-AA}
\end{figure}

For a single sheet of graphene including SOC, the tight binding Hamiltonian is\cite{Kane:2005}
\begin{equation}\label{H1SOC}
\hat{H}^0_{\tau_z,s_z}=\left(\begin{array}{cc}
\Delta\tau_z s_z & f(\bm{k})\\
f^*(\bm{k}) & -\Delta \tau_z s_z
\end{array}\right),
\end{equation}
where, in the continuum limit, $f(\bm{k})=\hbar v_F (\tau_z k_x-ik_y)$ and $\Delta=3\sqrt{3}t_{so}$,
with $t_{so}$, the next-nearest neighbour hopping amplitude. $\tau_z=\pm 1$ for the Dirac points ${K}$ and ${K'}$ 
and $s_z=\pm 1$ corresponding to the up/down spin component perpendicular to the graphene sheet\cite{Prada:2011}.  
With this, the Hamiltonian for the AA-stacked bilayer is
\begin{equation}\label{HAASOC}
\hat{H}_{\tau_z,s_z}=\left(\begin{array}{cc}
\hat H^0_{1\,\tau_z,s_z} & \hat H_\perp\\
\hat H_\perp & \hat H^0_{2\,\tau_z,s_z}
\end{array}\right),
\end{equation}
where we have used the eigenvector $\Psi=(a_{1\bm{k}}, b_{1 \bm{k}}, a_{2\bm{k}}, b_{2 \bm{k}})$.  
When dealing with the case of SOC in both layers, both $\hat H^0_{1\,\tau_z,s_z}$ and $\hat H^0_{2\,\tau_z,s_z}$ are given by Eqn.~\eqref{H1SOC} and 
$\hat H_\perp$ is the coupling between the layers, again taken to be 
\begin{equation}\label{HSOC}
\hat{H}_{\perp}=\left(\begin{array}{cc}
\gamma & 0\\
0 & \gamma
\end{array}\right)
\end{equation}
for AA-stacking.
Eqn.~\eqref{HAASOC} gives the four energy bands\cite{Prada:2011}
\begin{equation}\label{EnergySOC}
\varepsilon_\alpha(\bm{k})=\pm\left(\sqrt{\varepsilon^2+\Delta^2}+(-)^\alpha\gamma\right),
\end{equation}
where $\alpha=1$ and 2.  
These are illustrated in the inset of  Fig.~\ref{fig:Conductivity-SOC-AA}(b) for a 
particular $K$ point.

While we are interested in $\sigma_{xx}$, we will forgo providing the explicit details 
of the calculation as they can be developed by following the procedure already outlined earlier.  
The expression for $\sigma_{xx}(\Omega)$ given in Eqn.~\eqref{CondIntegral} can still be used.
However, in  lieu of the 
degeneracy factor $N_f$, a sum over $\tau_z$ and $s_z$ should be taken when using Eqn.~\eqref{HAASOC} to calculate the Green's function.  
Our velocity operator $\hat{v}_x$ written in the basis used for this section can be evaluated as before from
$\hbar\hat v_{x}=\partial \hat H/\partial k_x$.

The effect on the real part of the longitudinal conductivity of AA-stacked bilayer graphene when SOC is present in both layers can be seen in 
Fig.~\ref{fig:Conductivity-SOC-AA} for the case of (a) $\mu=0$ and (b) $\mu=0.5\gamma$.  The two shifted monolayer dispersions in the band structure
 are now gapped 
by $2\Delta$ about $\pm \gamma$ (see inset of Fig.~\ref{fig:Conductivity-SOC-AA}).  
For $\mu=0$ and $\Delta<\gamma$, as shown in
Fig.~\ref{fig:Conductivity-SOC-AA}(a), there is the usual Drude conductivity and a jump at $2\gamma$ as in the case of no SOC. However,
the shape of this jump is typical of a gapped electronic spectrum which gives rise to a discontinuity in the electronic density
of states. Indeed for the monolayer of graphene, such behavior has been calculated.\cite{Gusynin:2006}
As the frequency increases the usual bilayer background is recovered. The features of the conductivity show that, 
as in the case of no SOC, transitions are only allowed within each decoupled (like-colored) monolayer band.  For the case of $\Delta>\gamma$, we no 
longer have the Drude contribution as no states, and therefore no transitions, are available at zero energy.  The peak in the conductivity now occurs at $2\Delta$ and 
again we can see that we obtain the usual background conductivity for significantly high frequency. 
We have chosen our values for $\Delta/\gamma$ to be in keeping with the parameters used by Prada \emph{et al}.\cite{Prada:2011}, however, in graphene
the intrinsic SOC gap is $\sim 10^{-3}$ meV and hence $\Delta/\gamma\sim 10^{-5}$. Nonetheless, we have chosen this model
to indicate the effect of energy gaps appearing in the band structure. Indeed, other graphene-like systems are now being studied which have
much larger predicted SOC gaps, such as silicene ($\Delta\sim 1.5$ meV) and germanene ($\sim 25$ meV) which can be further
tuned by strain or perpendicular 
electric field.\cite{Liu:2011,Drummond:2012} Likewise, a varying mass gap of up to $\sim 150$ meV has been found in the 3D topological
insulator TlBi(S$_{\rm 1-x}$Se$_{\rm x}$)$_2$ by Sato \emph{et al}.\cite{Sato:2011} and the SOC gap in monolayer MoS$_2$ and other group-VI dichalcogenides
 is on the order of $1.5-1.8$ eV\cite{Xiao:2012}.
Consequently, the results of this section may be very relevant to future developments in these graphene-like systems.

In Fig.~\ref{fig:Conductivity-SOC-AA}(b), the effect of finite doping is considered in comparison with the case of
no SOC. For small enough $\Delta$, the SOC curve will track the one without SOC with the exception that there is a peak at each absorption
edge reflecting the energy gaps in the band structure at $k=0$. If $\Delta>|\gamma-\mu|$, the edge will be at $2\Delta$ rather than at
$2|\gamma-\mu|$, and likewise for $\Delta>\gamma+\mu$, there will be only one jump at $2\Delta$. A Drude contribution remains provided that $\Delta<\gamma+\mu$.
The behavior of the absorption with doping reflects possible transitions between the original shifted monolayer bands subject to the opening
of a gap. Mixing between these two monolayer-type bands does not occur. This is illustrated in the inset of Fig.~\ref{fig:Conductivity-SOC-AA}(b).
The behavior embodied by these figures can be derived  analytically and we find it to be:
\begin{equation}
\sigma_{xx}(\Omega)=\tilde{\sigma}_{xx}(\Omega,|\gamma-\mu|)+\tilde{\sigma}_{xx}(\Omega,\gamma+\mu),
\end{equation}
with
\begin{align}
\frac{\tilde{\sigma}_{xx}(\Omega,\Upsilon)}{\sigma_0}=&4\frac{(\Upsilon^2-\Delta^2)}{\Upsilon}\delta(\Omega)\Theta(\Upsilon-\Delta)\notag\\
&+\biggl[1+\biggl(\frac{2\Delta}{\Omega}\biggr)^2\biggr]\Theta[\Omega-2\, {\rm max}(\Upsilon,\Delta)],
\end{align}
where this last expression is the conductivity for massive Dirac quasiparticles\cite{Gusynin:2006,Tse:2010}.
This formula is in good agreement with the numerical work. Note that for the longitudinal optical conductivity, there is a conservation of spectral
weight upon introducing a finite $\mu$ and finite $\Delta$.  The corresponding imaginary conductivity is given by
\begin{equation}
\sigma_{xx}^{\prime\prime}(\Omega)=\tilde{\sigma}^{\prime\prime}_{xx}(\Omega,|\gamma-\mu|)+\tilde{\sigma}^{\prime\prime}_{xx}(\Omega,\gamma+\mu),
\end{equation}
where
\begin{align}
\frac{\tilde{\sigma}_{xx}^{\prime\prime}(\Omega,\Upsilon)}{\sigma_0}=&\frac{4}{\pi\Omega}\frac{\Upsilon^2-\Delta^2}{\Upsilon}\Theta(\Upsilon-\Delta)+\frac{1}{\pi}\text{ln}\bigg|\frac{\Omega-2\text{max}(\Upsilon,\Delta)}{\Omega+2\text{max}(\Upsilon,\Delta)}\bigg|\notag\\
&+\frac{4\Delta^2}{\pi\Omega}\bigg[\frac{1}{\text{max}(\Upsilon,\Delta)}+\frac{1}{\Omega}\text{ln}\bigg|\frac{\Omega-2\text{max}(\Upsilon,\Delta)}{\Omega+2\text{max}(\Upsilon,\Delta)}\bigg|\bigg].
\end{align}

\begin{figure}[h!]
\includegraphics[width=0.8\linewidth]{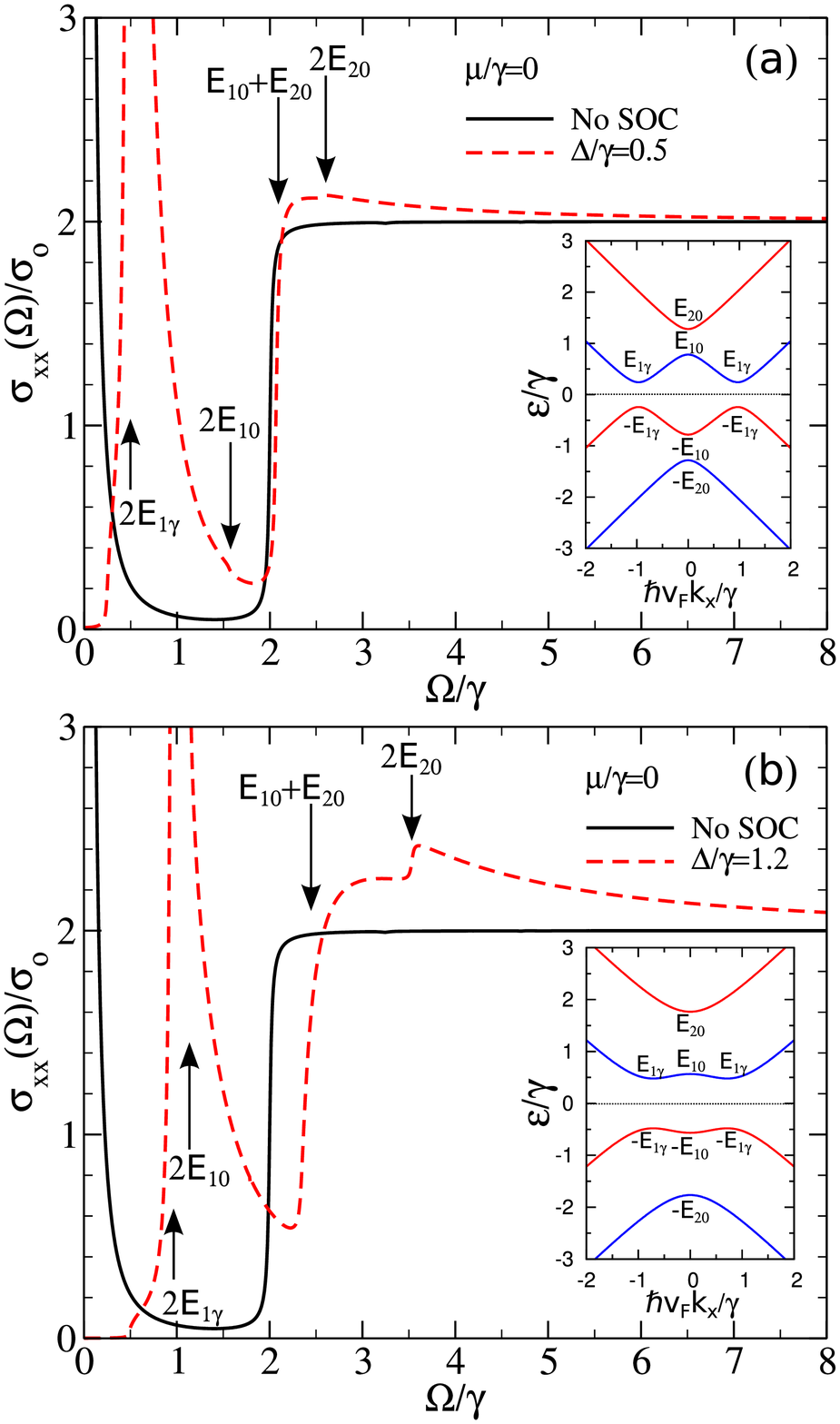}
\caption{(Color online)
Conductivity for the case of SOC in one layer only and the doping set to
charge neutrality. The value of $\Delta$ in the one
layer is (a) $\Delta=0.5\gamma$ and (b) $\Delta=1.2\gamma$. The band structure
for each case is shown as an inset.
}
\label{fig:Conductivity-SOC1-m0}
\end{figure}

\begin{figure}[h!]
\includegraphics[width=0.8\linewidth]{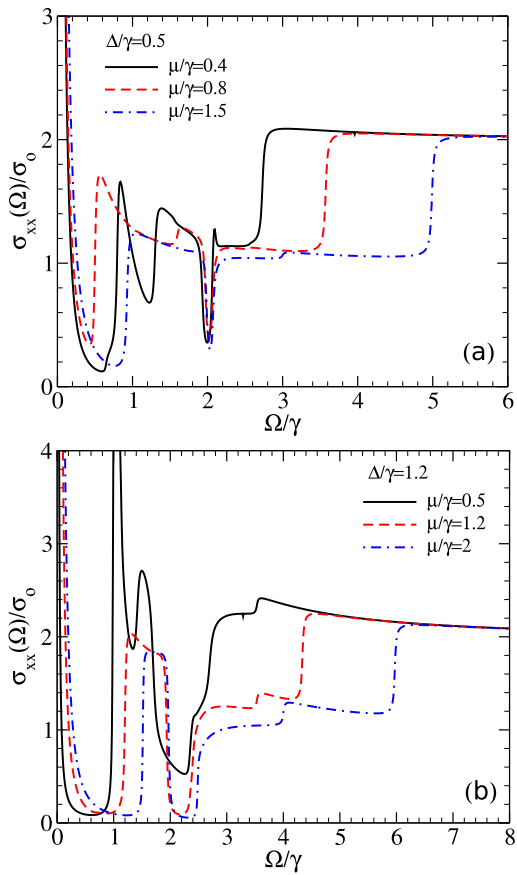}
\caption{(Color online)
As for Fig.~\ref{fig:Conductivity-SOC1-m0} but for
finite doping as indicated in the legends. The SOC is only in
one of the two layers with value (a) $\Delta=0.5\gamma$ 
and (b) $\Delta=1.2\gamma$.
}
\label{fig:Conductivity-SOC1-mu}
\end{figure}

If we only keep SOC in one layer, one of the diagonal elements of our Hamiltonian given by Eqn.~\eqref{HAASOC} becomes
\begin{equation}
\hat{H}_{\tau_z,s_z}=\left(\begin{array}{cc}
0 & f(\bm{k})\\
f^*(\bm{k}) & 0
\end{array}\right)
\end{equation}
and the four energy bands become\cite{Prada:2011}
\begin{equation}\label{EnergySOC1}
\varepsilon_\alpha(\bm{k})=\pm\sqrt{\varepsilon^2+\gamma^2+\frac{\Delta^2}{2}+(-)^\alpha\sqrt{4\varepsilon^2\gamma^2+\gamma^2\Delta^2+\frac{\Delta^4}{4}}},
\end{equation}
where $\alpha=1$ or 2. 
The conductivity for this case can be seen in Figs.~\ref{fig:Conductivity-SOC1-m0} and \ref{fig:Conductivity-SOC1-mu} 
for the case of $\mu=0$ and finite $\mu$, respectively.  The band structure is plotted in the insets of 
Fig.~\ref{fig:Conductivity-SOC1-m0}(a) and (b) for $\Delta=0.5\gamma$ and $\Delta=1.2\gamma$, respectively.
The key energy levels labelled in the insets are given by
\begin{equation} E_{\alpha\varepsilon}=\sqrt{\tilde\varepsilon^2+1+\frac{\tilde{\Delta}^2}{2}+
(-)^\alpha\sqrt{4\tilde\varepsilon^2+\tilde{\Delta}^2+\frac{\tilde{\Delta}^4}{4}}}, \end{equation}
where $\tilde{\Delta}=\Delta/\gamma$ and $\tilde\varepsilon=\varepsilon/\gamma$.  Gaps now appear about zero energy and about $\pm\gamma$.

The shape of the optical curves reveals that transitions are now allowed between every band unlike
in the previous case with SOC in  both layers.
In Fig.~\ref{fig:Conductivity-SOC1-m0}, a strong absorption is seen at $2E_{1\gamma}$ (see label in inset to
the figure) which is $\sim \Delta$ at small values of $\Delta$. The sharpness and strength of this
absorption feature is due to the existence of a square root singularity in the electronic
density of states at half this energy or $\sim\Delta/2$. This feature has replaced the Drude absorption
due to the gap about zero energy, but very little else changes in Fig.~\ref{fig:Conductivity-SOC1-m0}(a)
from the no-SOC case for small $\Delta<\gamma$. For $\Delta>\gamma$ shown in Fig.~\ref{fig:Conductivity-SOC1-m0}(b),
further structure appears which can be traced to various transitions in the band structure as has
been labelled in the inset. While the behavior is relatively simple for $\mu=0$, for finite $\mu$ it is much more
complicated. Fig.~\ref{fig:Conductivity-SOC1-mu}(a) and (b) shows the case for $\Delta=0.5\gamma$ and $\Delta=1.2\gamma$,
respectively, where $\mu$ is varied through key parts of the band structure [refer to the insets of Fig.~\ref{fig:Conductivity-SOC1-m0}].
While the high frequency behavior is similar to what we have seen for no SOC with an interband absorption
edge tracking $2(\gamma+\mu)$, the low frequency behavior is very structured. Notable is that the feature at $2E_{1\gamma}$ (or $\sim \Delta$
for small $\Delta$) is quickly suppressed by Pauli blocking at finite $\mu$ and there appears to be a deep absorption minimum
just on or after $\Omega=2\gamma$. Comparing this figure with Fig.~\ref{fig:Conductivity-AA} shows that
for each value of $\mu$ the underlying structure of the double step for the AA-stacked case with $\Delta=0$ is retained,
while the low frequency behavior oscillates about the $\Delta=0$ case. The governing behavior still appears to be dominated by
the decoupled monolayer bands however the complicated structure arising from various interband transitions within the different
bands is reminiscent of what is found in Bernal-stacked bilayer graphene  for finite $\mu$ and an asymmetry gap\cite{Nicol:2008} which gives
rise to similar looking (but not quite the same) band structure. We do not have an analytic form for this complex behavior  and one
must rely on an examination of the optical transitions available in the band structure to identify the detailed structure.
However, it is clear that the case of SOC in only one plane is quite different from that where SOC is in both planes.

\section{Conclusions}

We have examined the dynamical conductivity for AA-stacked bilayer graphene. The behavior is not simply
a case of doubling the conductivity of a graphene monolayer. Indeed the interlayer hopping $\gamma$ must appear
as an important energy scale. In contrast to the monolayer which is constant at all frequencies for charge
neutrality, the bilayer exhibits a Drude conductivity and absorption is Pauli-blocked for frequencies less
than $2\gamma$. At finite doping, a double step occurs in the interband absorption with onset for each
step at $|\gamma-\mu|$ and $\gamma+\mu$ and the behavior is non-monotonic in $\mu$ as a result. The perpendicular
response is completely centered on $2\gamma$ with the absorption peak being very significant at all dopings
and increasing with doping. This is unlike the behavior seen in AB-stacked bilayer graphene. Applying a bias across the bilayer does not open an energy gap in the band structure but merely renormalizes the effective interlayer hopping to a greater value.  The conductivity of the trilayer exhibits three absorption edges leading to a flat conductivity background which steps up through $\sigma_0$ to $3\sigma_0$.  The conductivity in this case is clearly seen to be a sum of that due to a monolayer added to that of a bilayer with interlayer hopping $\sqrt{2}\gamma$.  Including spin orbit
coupling in each layer of the AA-stacked bilayer leads to double step behavior where the absorption edges show a peak due to a gap of $2\Delta$ in the
band structure and the location is set by an interplay between the energy scales of chemical potential $\mu$, $\gamma$
and $\Delta$. With spin orbit coupling in only one plane, an absorption peak is found at about $\Delta$ and very complicated
structure as a function of frequency results at finite $\mu$  although there remains a remnant of the
underlying structure of simple AA-stacked bilayer conductivity. As experimental isolation of AA-stacked graphene has been reported
and interest in topological insulators with spin orbit coupling is high, this study may be timely for future work in these
areas.

\begin{acknowledgments}
We thank James LeBlanc, Jules Carbotte, and Bernie Nickel
 for  helpful discussions.
This work has been supported by NSERC of Canada
and in part by the National Science Foundation
 under Grant No. NSF PHY11-25915.  
\end{acknowledgments}

\bibliographystyle{apsrev4-1}
\bibliography{aabi-bib}

\begin{thebibliography}{65}%
\makeatletter
\providecommand \@ifxundefined [1]{%
 \@ifx{#1\undefined}
}%
\providecommand \@ifnum [1]{%
 \ifnum #1\expandafter \@firstoftwo
 \else \expandafter \@secondoftwo
 \fi
}%
\providecommand \@ifx [1]{%
 \ifx #1\expandafter \@firstoftwo
 \else \expandafter \@secondoftwo
 \fi
}%
\providecommand \natexlab [1]{#1}%
\providecommand \enquote  [1]{``#1''}%
\providecommand \bibnamefont  [1]{#1}%
\providecommand \bibfnamefont [1]{#1}%
\providecommand \citenamefont [1]{#1}%
\providecommand \href@noop [0]{\@secondoftwo}%
\providecommand \href [0]{\begingroup \@sanitize@url \@href}%
\providecommand \@href[1]{\@@startlink{#1}\@@href}%
\providecommand \@@href[1]{\endgroup#1\@@endlink}%
\providecommand \@sanitize@url [0]{\catcode `\\12\catcode `\$12\catcode
  `\&12\catcode `\#12\catcode `\^12\catcode `\_12\catcode `\%12\relax}%
\providecommand \@@startlink[1]{}%
\providecommand \@@endlink[0]{}%
\providecommand \url  [0]{\begingroup\@sanitize@url \@url }%
\providecommand \@url [1]{\endgroup\@href {#1}{\urlprefix }}%
\providecommand \urlprefix  [0]{URL }%
\providecommand \Eprint [0]{\href }%
\providecommand \doibase [0]{http://dx.doi.org/}%
\providecommand \selectlanguage [0]{\@gobble}%
\providecommand \bibinfo  [0]{\@secondoftwo}%
\providecommand \bibfield  [0]{\@secondoftwo}%
\providecommand \translation [1]{[#1]}%
\providecommand \BibitemOpen [0]{}%
\providecommand \bibitemStop [0]{}%
\providecommand \bibitemNoStop [0]{.\EOS\space}%
\providecommand \EOS [0]{\spacefactor3000\relax}%
\providecommand \BibitemShut  [1]{\csname bibitem#1\endcsname}%
\let\auto@bib@innerbib\@empty
\bibitem [{\citenamefont {Wallace}(1947)}]{Wallace:1947}%
  \BibitemOpen
  \bibfield  {author} {\bibinfo {author} {\bibfnamefont {P.~R.}\ \bibnamefont
  {Wallace}},\ }\href@noop {} {\bibfield  {journal} {\bibinfo  {journal} {Phys.
  Rev.}\ }\textbf {\bibinfo {volume} {71}},\ \bibinfo {pages} {622} (\bibinfo
  {year} {1947})}\BibitemShut {NoStop}%
\bibitem [{\citenamefont {Semenoff}(1984)}]{Semenoff:1984}%
  \BibitemOpen
  \bibfield  {author} {\bibinfo {author} {\bibfnamefont {G.~W.}\ \bibnamefont
  {Semenoff}},\ }\href@noop {} {\bibfield  {journal} {\bibinfo  {journal}
  {Phys. Rev. Lett.}\ }\textbf {\bibinfo {volume} {53}},\ \bibinfo {pages}
  {2449} (\bibinfo {year} {1984})}\BibitemShut {NoStop}%
\bibitem [{\citenamefont {Novoselov}\ \emph {et~al.}(2004)\citenamefont
  {Novoselov}, \citenamefont {Geim}, \citenamefont {Morozov}, \citenamefont
  {Jiang}, \citenamefont {Zhang}, \citenamefont {Dubonos}, \citenamefont
  {Grigorieva},\ and\ \citenamefont {Firsov}}]{Novoselov:2004}%
  \BibitemOpen
  \bibfield  {author} {\bibinfo {author} {\bibfnamefont {K.~S.}\ \bibnamefont
  {Novoselov}}, \bibinfo {author} {\bibfnamefont {A.~K.}\ \bibnamefont {Geim}},
  \bibinfo {author} {\bibfnamefont {S.~V.}\ \bibnamefont {Morozov}}, \bibinfo
  {author} {\bibfnamefont {D.}~\bibnamefont {Jiang}}, \bibinfo {author}
  {\bibfnamefont {Y.}~\bibnamefont {Zhang}}, \bibinfo {author} {\bibfnamefont
  {S.~V.}\ \bibnamefont {Dubonos}}, \bibinfo {author} {\bibfnamefont {I.~V.}\
  \bibnamefont {Grigorieva}}, \ and\ \bibinfo {author} {\bibfnamefont {A.~A.}\
  \bibnamefont {Firsov}},\ }\href@noop {} {\bibfield  {journal} {\bibinfo
  {journal} {Science}\ }\textbf {\bibinfo {volume} {306}},\ \bibinfo {pages}
  {666} (\bibinfo {year} {2004})}\BibitemShut {NoStop}%
\bibitem [{\citenamefont {Novoselov}\ \emph
  {et~al.}(2005{\natexlab{a}})\citenamefont {Novoselov}, \citenamefont {Jiang},
  \citenamefont {Schedin}, \citenamefont {Booth}, \citenamefont {Khotkevich},
  \citenamefont {Morozov},\ and\ \citenamefont {Geim}}]{Novoselov:2005a}%
  \BibitemOpen
  \bibfield  {author} {\bibinfo {author} {\bibfnamefont {K.~S.}\ \bibnamefont
  {Novoselov}}, \bibinfo {author} {\bibfnamefont {D.}~\bibnamefont {Jiang}},
  \bibinfo {author} {\bibfnamefont {F.}~\bibnamefont {Schedin}}, \bibinfo
  {author} {\bibfnamefont {T.~J.}\ \bibnamefont {Booth}}, \bibinfo {author}
  {\bibfnamefont {V.~V.}\ \bibnamefont {Khotkevich}}, \bibinfo {author}
  {\bibfnamefont {S.~V.}\ \bibnamefont {Morozov}}, \ and\ \bibinfo {author}
  {\bibfnamefont {A.~K.}\ \bibnamefont {Geim}},\ }\href@noop {} {\bibfield
  {journal} {\bibinfo  {journal} {Proc. Natl Acad. Sci. USA.}\ }\textbf
  {\bibinfo {volume} {102}},\ \bibinfo {pages} {10451} (\bibinfo {year}
  {2005}{\natexlab{a}})}\BibitemShut {NoStop}%
\bibitem [{\citenamefont {Zhang}\ \emph {et~al.}(2005)\citenamefont {Zhang},
  \citenamefont {Tan}, \citenamefont {Stormer},\ and\ \citenamefont
  {Kim}}]{Zhang:2005}%
  \BibitemOpen
  \bibfield  {author} {\bibinfo {author} {\bibfnamefont {Y.}~\bibnamefont
  {Zhang}}, \bibinfo {author} {\bibfnamefont {Y.-W.}\ \bibnamefont {Tan}},
  \bibinfo {author} {\bibfnamefont {H.~L.}\ \bibnamefont {Stormer}}, \ and\
  \bibinfo {author} {\bibfnamefont {P.}~\bibnamefont {Kim}},\ }\href@noop {}
  {\bibfield  {journal} {\bibinfo  {journal} {Nature}\ }\textbf {\bibinfo
  {volume} {438}},\ \bibinfo {pages} {201} (\bibinfo {year}
  {2005})}\BibitemShut {NoStop}%
\bibitem [{\citenamefont {Novoselov}\ \emph
  {et~al.}(2005{\natexlab{b}})\citenamefont {Novoselov}, \citenamefont {Geim},
  \citenamefont {Morozov}, \citenamefont {Jiang}, \citenamefont {Katsnelson},
  \citenamefont {Grigorieva}, \citenamefont {Dubonos},\ and\ \citenamefont
  {Firsov}}]{Novoselov:2005b}%
  \BibitemOpen
  \bibfield  {author} {\bibinfo {author} {\bibfnamefont {K.~S.}\ \bibnamefont
  {Novoselov}}, \bibinfo {author} {\bibfnamefont {A.~K.}\ \bibnamefont {Geim}},
  \bibinfo {author} {\bibfnamefont {S.~V.}\ \bibnamefont {Morozov}}, \bibinfo
  {author} {\bibfnamefont {D.}~\bibnamefont {Jiang}}, \bibinfo {author}
  {\bibfnamefont {M.~I.}\ \bibnamefont {Katsnelson}}, \bibinfo {author}
  {\bibfnamefont {I.~V.}\ \bibnamefont {Grigorieva}}, \bibinfo {author}
  {\bibfnamefont {S.~V.}\ \bibnamefont {Dubonos}}, \ and\ \bibinfo {author}
  {\bibfnamefont {A.~A.}\ \bibnamefont {Firsov}},\ }\href@noop {} {\bibfield
  {journal} {\bibinfo  {journal} {Nature}\ }\textbf {\bibinfo {volume} {438}},\
  \bibinfo {pages} {197} (\bibinfo {year} {2005}{\natexlab{b}})}\BibitemShut
  {NoStop}%
\bibitem [{\citenamefont {Crassee}\ \emph {et~al.}(2011)\citenamefont
  {Crassee}, \citenamefont {Levallois}, \citenamefont {Walter}, \citenamefont
  {Ostler}, \citenamefont {Bostwick}, \citenamefont {Rotenberg}, \citenamefont
  {Seyller}, \citenamefont {van~der Marel},\ and\ \citenamefont
  {Kuzmenko}}]{Crassee:2011}%
  \BibitemOpen
  \bibfield  {author} {\bibinfo {author} {\bibfnamefont {I.}~\bibnamefont
  {Crassee}}, \bibinfo {author} {\bibfnamefont {J.}~\bibnamefont {Levallois}},
  \bibinfo {author} {\bibfnamefont {A.~L.}\ \bibnamefont {Walter}}, \bibinfo
  {author} {\bibfnamefont {M.}~\bibnamefont {Ostler}}, \bibinfo {author}
  {\bibfnamefont {A.}~\bibnamefont {Bostwick}}, \bibinfo {author}
  {\bibfnamefont {E.}~\bibnamefont {Rotenberg}}, \bibinfo {author}
  {\bibfnamefont {T.}~\bibnamefont {Seyller}}, \bibinfo {author} {\bibfnamefont
  {D.}~\bibnamefont {van~der Marel}}, \ and\ \bibinfo {author} {\bibfnamefont
  {A.~B.}\ \bibnamefont {Kuzmenko}},\ }\href@noop {} {\bibfield  {journal}
  {\bibinfo  {journal} {Nature Phys.}\ }\textbf {\bibinfo {volume} {7}},\
  \bibinfo {pages} {48} (\bibinfo {year} {2011})}\BibitemShut {NoStop}%
\bibitem [{\citenamefont {Bostwick}\ \emph {et~al.}(2010)\citenamefont
  {Bostwick}, \citenamefont {Speck}, \citenamefont {Seyller}, \citenamefont
  {Horn}, \citenamefont {Polini}, \citenamefont {Asgari}, \citenamefont
  {MacDonald},\ and\ \citenamefont {Rotenberg}}]{Bostwick:2010}%
  \BibitemOpen
  \bibfield  {author} {\bibinfo {author} {\bibfnamefont {A.}~\bibnamefont
  {Bostwick}}, \bibinfo {author} {\bibfnamefont {F.}~\bibnamefont {Speck}},
  \bibinfo {author} {\bibfnamefont {T.}~\bibnamefont {Seyller}}, \bibinfo
  {author} {\bibfnamefont {K.}~\bibnamefont {Horn}}, \bibinfo {author}
  {\bibfnamefont {M.}~\bibnamefont {Polini}}, \bibinfo {author} {\bibfnamefont
  {R.}~\bibnamefont {Asgari}}, \bibinfo {author} {\bibfnamefont {A.~H.}\
  \bibnamefont {MacDonald}}, \ and\ \bibinfo {author} {\bibfnamefont
  {E.}~\bibnamefont {Rotenberg}},\ }\href@noop {} {\bibfield  {journal}
  {\bibinfo  {journal} {Science}\ }\textbf {\bibinfo {volume} {328}},\ \bibinfo
  {pages} {999} (\bibinfo {year} {2010})}\BibitemShut {NoStop}%
\bibitem [{\citenamefont {{Castro Neto}}\ \emph {et~al.}(2009)\citenamefont
  {{Castro Neto}}, \citenamefont {Guinea}, \citenamefont {Peres}, \citenamefont
  {Novoselov},\ and\ \citenamefont {Geim}}]{Neto:2009}%
  \BibitemOpen
  \bibfield  {author} {\bibinfo {author} {\bibfnamefont {A.~H.}\ \bibnamefont
  {{Castro Neto}}}, \bibinfo {author} {\bibfnamefont {F.}~\bibnamefont
  {Guinea}}, \bibinfo {author} {\bibfnamefont {N.~M.~R.}\ \bibnamefont
  {Peres}}, \bibinfo {author} {\bibfnamefont {K.~S.}\ \bibnamefont
  {Novoselov}}, \ and\ \bibinfo {author} {\bibfnamefont {A.~K.}\ \bibnamefont
  {Geim}},\ }\href@noop {} {\bibfield  {journal} {\bibinfo  {journal} {Rev.
  Mod. Phys.}\ }\textbf {\bibinfo {volume} {81}},\ \bibinfo {pages} {109}
  (\bibinfo {year} {2009})}\BibitemShut {NoStop}%
\bibitem [{\citenamefont {Abergel}\ \emph {et~al.}(2010)\citenamefont
  {Abergel}, \citenamefont {Apalkov}, \citenamefont {Berashevich},
  \citenamefont {Ziegler},\ and\ \citenamefont {Chakraborty}}]{Abergel:2010}%
  \BibitemOpen
  \bibfield  {author} {\bibinfo {author} {\bibfnamefont {D.~S.~L.}\
  \bibnamefont {Abergel}}, \bibinfo {author} {\bibfnamefont {V.}~\bibnamefont
  {Apalkov}}, \bibinfo {author} {\bibfnamefont {J.}~\bibnamefont
  {Berashevich}}, \bibinfo {author} {\bibfnamefont {K.}~\bibnamefont
  {Ziegler}}, \ and\ \bibinfo {author} {\bibfnamefont {T.}~\bibnamefont
  {Chakraborty}},\ }\href@noop {} {\bibfield  {journal} {\bibinfo  {journal}
  {Adv. in Phys.}\ }\textbf {\bibinfo {volume} {59}},\ \bibinfo {pages} {261}
  (\bibinfo {year} {2010})}\BibitemShut {NoStop}%
\bibitem [{\citenamefont {Das~Sarma}\ \emph {et~al.}(2011)\citenamefont
  {Das~Sarma}, \citenamefont {Adam}, \citenamefont {Hwang},\ and\ \citenamefont
  {Rossi}}]{DasSarma:2011}%
  \BibitemOpen
  \bibfield  {author} {\bibinfo {author} {\bibfnamefont {S.}~\bibnamefont
  {Das~Sarma}}, \bibinfo {author} {\bibfnamefont {S.}~\bibnamefont {Adam}},
  \bibinfo {author} {\bibfnamefont {E.~H.}\ \bibnamefont {Hwang}}, \ and\
  \bibinfo {author} {\bibfnamefont {E.}~\bibnamefont {Rossi}},\ }\href@noop {}
  {\bibfield  {journal} {\bibinfo  {journal} {Rev. Mod. Phys.}\ }\textbf
  {\bibinfo {volume} {83}},\ \bibinfo {pages} {407} (\bibinfo {year}
  {2011})}\BibitemShut {NoStop}%
\bibitem [{\citenamefont {Kotov}\ \emph {et~al.}(2010)\citenamefont {Kotov},
  \citenamefont {Uchoa}, \citenamefont {Pereira}, \citenamefont {Guinea},\ and\
  \citenamefont {{Castro Neto}}}]{Kotov:2010}%
  \BibitemOpen
  \bibfield  {author} {\bibinfo {author} {\bibfnamefont {V.~N.}\ \bibnamefont
  {Kotov}}, \bibinfo {author} {\bibfnamefont {B.}~\bibnamefont {Uchoa}},
  \bibinfo {author} {\bibfnamefont {V.~M.}\ \bibnamefont {Pereira}}, \bibinfo
  {author} {\bibfnamefont {F.}~\bibnamefont {Guinea}}, \ and\ \bibinfo {author}
  {\bibfnamefont {A.~H.}\ \bibnamefont {{Castro Neto}}},\ }\href@noop {} {}
  (\bibinfo {year} {2010}),\ \Eprint {http://arxiv.org/abs/arXiv:1012.3484v2}
  {arXiv:1012.3484v2} \BibitemShut {NoStop}%
\bibitem [{\citenamefont {Geim}\ and\ \citenamefont
  {Novoselov}(2007)}]{Geim:2007}%
  \BibitemOpen
  \bibfield  {author} {\bibinfo {author} {\bibfnamefont {A.~K.}\ \bibnamefont
  {Geim}}\ and\ \bibinfo {author} {\bibfnamefont {K.~S.}\ \bibnamefont
  {Novoselov}},\ }\href@noop {} {\bibfield  {journal} {\bibinfo  {journal}
  {Nature Materials}\ }\textbf {\bibinfo {volume} {6}},\ \bibinfo {pages} {183}
  (\bibinfo {year} {2007})}\BibitemShut {NoStop}%
\bibitem [{\citenamefont {McCann}\ and\ \citenamefont
  {Fal'ko}(2006)}]{McCann:2006a}%
  \BibitemOpen
  \bibfield  {author} {\bibinfo {author} {\bibfnamefont {E.}~\bibnamefont
  {McCann}}\ and\ \bibinfo {author} {\bibfnamefont {V.~I.}\ \bibnamefont
  {Fal'ko}},\ }\href@noop {} {\bibfield  {journal} {\bibinfo  {journal} {Phys.
  Rev. Lett.}\ }\textbf {\bibinfo {volume} {96}},\ \bibinfo {pages} {086805}
  (\bibinfo {year} {2006})}\BibitemShut {NoStop}%
\bibitem [{\citenamefont {McCann}\ and\ \citenamefont
  {Koshino}(2012)}]{McCann:2012}%
  \BibitemOpen
  \bibfield  {author} {\bibinfo {author} {\bibfnamefont {E.}~\bibnamefont
  {McCann}}\ and\ \bibinfo {author} {\bibfnamefont {M.}~\bibnamefont
  {Koshino}},\ }\href@noop {} {} (\bibinfo {year} {2012}),\ \Eprint
  {http://arxiv.org/abs/arXiv:1205.6953v1} {arXiv:1205.6953v1} \BibitemShut
  {NoStop}%
\bibitem [{\citenamefont {Ohta}\ \emph {et~al.}(2006)\citenamefont {Ohta},
  \citenamefont {Bostwick}, \citenamefont {Seyller}, \citenamefont {Horn},\
  and\ \citenamefont {Rotenberg}}]{Ohta:2006}%
  \BibitemOpen
  \bibfield  {author} {\bibinfo {author} {\bibfnamefont {T.}~\bibnamefont
  {Ohta}}, \bibinfo {author} {\bibfnamefont {A.}~\bibnamefont {Bostwick}},
  \bibinfo {author} {\bibfnamefont {T.}~\bibnamefont {Seyller}}, \bibinfo
  {author} {\bibfnamefont {K.}~\bibnamefont {Horn}}, \ and\ \bibinfo {author}
  {\bibfnamefont {E.}~\bibnamefont {Rotenberg}},\ }\href@noop {} {\bibfield
  {journal} {\bibinfo  {journal} {Science}\ }\textbf {\bibinfo {volume}
  {313}},\ \bibinfo {pages} {951} (\bibinfo {year} {2006})}\BibitemShut
  {NoStop}%
\bibitem [{\citenamefont {Li}\ \emph {et~al.}(2010)\citenamefont {Li},
  \citenamefont {Luican}, \citenamefont {dos Santos}, \citenamefont {{Castro
  Neto}}, \citenamefont {Reina}, \citenamefont {Kong},\ and\ \citenamefont
  {Andrei}}]{Li:2010}%
  \BibitemOpen
  \bibfield  {author} {\bibinfo {author} {\bibfnamefont {G.}~\bibnamefont
  {Li}}, \bibinfo {author} {\bibfnamefont {A.}~\bibnamefont {Luican}}, \bibinfo
  {author} {\bibfnamefont {J.~M. B.~L.}\ \bibnamefont {dos Santos}}, \bibinfo
  {author} {\bibfnamefont {A.~H.}\ \bibnamefont {{Castro Neto}}}, \bibinfo
  {author} {\bibfnamefont {A.}~\bibnamefont {Reina}}, \bibinfo {author}
  {\bibfnamefont {J.}~\bibnamefont {Kong}}, \ and\ \bibinfo {author}
  {\bibfnamefont {E.~Y.}\ \bibnamefont {Andrei}},\ }\href@noop {} {\bibfield
  {journal} {\bibinfo  {journal} {Nature Phys.}\ }\textbf {\bibinfo {volume}
  {6}},\ \bibinfo {pages} {109} (\bibinfo {year} {2010})}\BibitemShut {NoStop}%
\bibitem [{\citenamefont {Luican}\ \emph {et~al.}(2011)\citenamefont {Luican},
  \citenamefont {Li}, \citenamefont {Reina}, \citenamefont {Kong},
  \citenamefont {Nair}, \citenamefont {Novoselov}, \citenamefont {Geim},\ and\
  \citenamefont {Andrei}}]{Luican:2011}%
  \BibitemOpen
  \bibfield  {author} {\bibinfo {author} {\bibfnamefont {A.}~\bibnamefont
  {Luican}}, \bibinfo {author} {\bibfnamefont {G.}~\bibnamefont {Li}}, \bibinfo
  {author} {\bibfnamefont {A.}~\bibnamefont {Reina}}, \bibinfo {author}
  {\bibfnamefont {J.}~\bibnamefont {Kong}}, \bibinfo {author} {\bibfnamefont
  {R.~R.}\ \bibnamefont {Nair}}, \bibinfo {author} {\bibfnamefont {K.~S.}\
  \bibnamefont {Novoselov}}, \bibinfo {author} {\bibfnamefont {A.~K.}\
  \bibnamefont {Geim}}, \ and\ \bibinfo {author} {\bibfnamefont {E.~Y.}\
  \bibnamefont {Andrei}},\ }\href@noop {} {\bibfield  {journal} {\bibinfo
  {journal} {Phys. Rev. Lett.}\ }\textbf {\bibinfo {volume} {106}},\ \bibinfo
  {pages} {126802} (\bibinfo {year} {2011})}\BibitemShut {NoStop}%
\bibitem [{\citenamefont {de~Laissardière}\ \emph {et~al.}(2010)\citenamefont
  {de~Laissardière}, \citenamefont {Mayou},\ and\ \citenamefont
  {Magaud}}]{Trambly:2010}%
  \BibitemOpen
  \bibfield  {author} {\bibinfo {author} {\bibfnamefont {G.~T.}\ \bibnamefont
  {de~Laissardière}}, \bibinfo {author} {\bibfnamefont {D.}~\bibnamefont
  {Mayou}}, \ and\ \bibinfo {author} {\bibfnamefont {L.}~\bibnamefont
  {Magaud}},\ }\href@noop {} {\bibfield  {journal} {\bibinfo  {journal} {Nano
  Letters}\ }\textbf {\bibinfo {volume} {10}},\ \bibinfo {pages} {804}
  (\bibinfo {year} {2010})}\BibitemShut {NoStop}%
\bibitem [{\citenamefont {de~Laissardière}\ \emph {et~al.}(2012)\citenamefont
  {de~Laissardière}, \citenamefont {Mayou},\ and\ \citenamefont
  {Magaud}}]{Trambly:arxiv}%
  \BibitemOpen
  \bibfield  {author} {\bibinfo {author} {\bibfnamefont {G.~T.}\ \bibnamefont
  {de~Laissardière}}, \bibinfo {author} {\bibfnamefont {D.}~\bibnamefont
  {Mayou}}, \ and\ \bibinfo {author} {\bibfnamefont {L.}~\bibnamefont
  {Magaud}},\ }\href@noop {} {} (\bibinfo {year} {2012}),\ \Eprint
  {http://arxiv.org/abs/arXiv:1203.3144v1} {arXiv:1203.3144v1} \BibitemShut
  {NoStop}%
\bibitem [{\citenamefont {Liu}\ \emph {et~al.}(2009)\citenamefont {Liu},
  \citenamefont {Suenaga}, \citenamefont {Harris},\ and\ \citenamefont
  {Iijima}}]{Liu:2009}%
  \BibitemOpen
  \bibfield  {author} {\bibinfo {author} {\bibfnamefont {Z.}~\bibnamefont
  {Liu}}, \bibinfo {author} {\bibfnamefont {K.}~\bibnamefont {Suenaga}},
  \bibinfo {author} {\bibfnamefont {P.~J.~F.}\ \bibnamefont {Harris}}, \ and\
  \bibinfo {author} {\bibfnamefont {S.}~\bibnamefont {Iijima}},\ }\href@noop {}
  {\bibfield  {journal} {\bibinfo  {journal} {Phys. Rev. Lett.}\ }\textbf
  {\bibinfo {volume} {102}},\ \bibinfo {pages} {015501} (\bibinfo {year}
  {2009})}\BibitemShut {NoStop}%
\bibitem [{\citenamefont {Borysiuk}\ \emph {et~al.}(2011)\citenamefont
  {Borysiuk}, \citenamefont {Soltys},\ and\ \citenamefont
  {Piechota}}]{Borysiuk:2011}%
  \BibitemOpen
  \bibfield  {author} {\bibinfo {author} {\bibfnamefont {J.}~\bibnamefont
  {Borysiuk}}, \bibinfo {author} {\bibfnamefont {J.}~\bibnamefont {Soltys}}, \
  and\ \bibinfo {author} {\bibfnamefont {J.}~\bibnamefont {Piechota}},\
  }\href@noop {} {\bibfield  {journal} {\bibinfo  {journal} {J. Appl. Phys.}\
  }\textbf {\bibinfo {volume} {109}},\ \bibinfo {pages} {093523} (\bibinfo
  {year} {2011})}\BibitemShut {NoStop}%
\bibitem [{\citenamefont {Rakhmanov}\ \emph {et~al.}(2011)\citenamefont
  {Rakhmanov}, \citenamefont {Rozhkov}, \citenamefont {Sboychakov},\ and\
  \citenamefont {Nori}}]{Rakhmanov:2011}%
  \BibitemOpen
  \bibfield  {author} {\bibinfo {author} {\bibfnamefont {A.~L.}\ \bibnamefont
  {Rakhmanov}}, \bibinfo {author} {\bibfnamefont {A.~V.}\ \bibnamefont
  {Rozhkov}}, \bibinfo {author} {\bibfnamefont {A.~O.}\ \bibnamefont
  {Sboychakov}}, \ and\ \bibinfo {author} {\bibfnamefont {F.}~\bibnamefont
  {Nori}},\ }\href@noop {} {} (\bibinfo {year} {2011}),\ \Eprint
  {http://arxiv.org/abs/arXiv:1111.5093v1} {arXiv:1111.5093v1} \BibitemShut
  {NoStop}%
\bibitem [{\citenamefont {Ando}\ \emph {et~al.}(2002)\citenamefont {Ando},
  \citenamefont {Zheng},\ and\ \citenamefont {Suzuura}}]{Ando:2002}%
  \BibitemOpen
  \bibfield  {author} {\bibinfo {author} {\bibfnamefont {T.}~\bibnamefont
  {Ando}}, \bibinfo {author} {\bibfnamefont {Y.}~\bibnamefont {Zheng}}, \ and\
  \bibinfo {author} {\bibfnamefont {H.}~\bibnamefont {Suzuura}},\ }\href@noop
  {} {\bibfield  {journal} {\bibinfo  {journal} {J. Phys. Soc. Jpn.}\ }\textbf
  {\bibinfo {volume} {71}},\ \bibinfo {pages} {1318} (\bibinfo {year}
  {2002})}\BibitemShut {NoStop}%
\bibitem [{\citenamefont {Gusynin}\ and\ \citenamefont
  {Sharapov}(2006)}]{Gusynin:2006a}%
  \BibitemOpen
  \bibfield  {author} {\bibinfo {author} {\bibfnamefont {V.~P.}\ \bibnamefont
  {Gusynin}}\ and\ \bibinfo {author} {\bibfnamefont {S.~G.}\ \bibnamefont
  {Sharapov}},\ }\href@noop {} {\bibfield  {journal} {\bibinfo  {journal}
  {Phys. Rev. B}\ }\textbf {\bibinfo {volume} {73}},\ \bibinfo {pages} {245411}
  (\bibinfo {year} {2006})}\BibitemShut {NoStop}%
\bibitem [{\citenamefont {Gusynin}\ \emph {et~al.}(2006)\citenamefont
  {Gusynin}, \citenamefont {Sharapov},\ and\ \citenamefont
  {Carbotte}}]{Gusynin:2006}%
  \BibitemOpen
  \bibfield  {author} {\bibinfo {author} {\bibfnamefont {V.~P.}\ \bibnamefont
  {Gusynin}}, \bibinfo {author} {\bibfnamefont {S.~G.}\ \bibnamefont
  {Sharapov}}, \ and\ \bibinfo {author} {\bibfnamefont {J.~P.}\ \bibnamefont
  {Carbotte}},\ }\href@noop {} {\bibfield  {journal} {\bibinfo  {journal}
  {Phys. Rev. Lett.}\ }\textbf {\bibinfo {volume} {96}},\ \bibinfo {pages}
  {256802} (\bibinfo {year} {2006})}\BibitemShut {NoStop}%
\bibitem [{\citenamefont {Falkovsky}\ and\ \citenamefont
  {Varlamov}(2007)}]{Falkovsky:2007}%
  \BibitemOpen
  \bibfield  {author} {\bibinfo {author} {\bibfnamefont {L.~A.}\ \bibnamefont
  {Falkovsky}}\ and\ \bibinfo {author} {\bibfnamefont {A.~A.}\ \bibnamefont
  {Varlamov}},\ }\href@noop {} {\bibfield  {journal} {\bibinfo  {journal} {Eur.
  Phys. J. B}\ }\textbf {\bibinfo {volume} {56}},\ \bibinfo {pages} {281}
  (\bibinfo {year} {2007})}\BibitemShut {NoStop}%
\bibitem [{\citenamefont {Peres}\ \emph {et~al.}(2006)\citenamefont {Peres},
  \citenamefont {Guinea},\ and\ \citenamefont {Castro~Neto}}]{Peres:2006}%
  \BibitemOpen
  \bibfield  {author} {\bibinfo {author} {\bibfnamefont {N.~M.~R.}\
  \bibnamefont {Peres}}, \bibinfo {author} {\bibfnamefont {F.}~\bibnamefont
  {Guinea}}, \ and\ \bibinfo {author} {\bibfnamefont {A.~H.}\ \bibnamefont
  {Castro~Neto}},\ }\href {\doibase 10.1103/PhysRevB.73.125411} {\bibfield
  {journal} {\bibinfo  {journal} {Phys. Rev. B}\ }\textbf {\bibinfo {volume}
  {73}},\ \bibinfo {pages} {125411} (\bibinfo {year} {2006})}\BibitemShut
  {NoStop}%
\bibitem [{\citenamefont {Gusynin}\ \emph {et~al.}(2007)\citenamefont
  {Gusynin}, \citenamefont {Sharapov},\ and\ \citenamefont
  {Carbotte}}]{Gusynin:2007}%
  \BibitemOpen
  \bibfield  {author} {\bibinfo {author} {\bibfnamefont {V.~P.}\ \bibnamefont
  {Gusynin}}, \bibinfo {author} {\bibfnamefont {S.~G.}\ \bibnamefont
  {Sharapov}}, \ and\ \bibinfo {author} {\bibfnamefont {J.~P.}\ \bibnamefont
  {Carbotte}},\ }\href@noop {} {\bibfield  {journal} {\bibinfo  {journal} {Int.
  J. Mod. Phys. B}\ }\textbf {\bibinfo {volume} {21}},\ \bibinfo {pages} {4611}
  (\bibinfo {year} {2007})}\BibitemShut {NoStop}%
\bibitem [{\citenamefont {Li}\ \emph {et~al.}(2008)\citenamefont {Li},
  \citenamefont {Henriksen}, \citenamefont {Jiang}, \citenamefont {Hao},
  \citenamefont {Martin}, \citenamefont {Kim}, \citenamefont {Stormer},\ and\
  \citenamefont {Basov}}]{Li:2008}%
  \BibitemOpen
  \bibfield  {author} {\bibinfo {author} {\bibfnamefont {Z.~Q.}\ \bibnamefont
  {Li}}, \bibinfo {author} {\bibfnamefont {E.~A.}\ \bibnamefont {Henriksen}},
  \bibinfo {author} {\bibfnamefont {Z.}~\bibnamefont {Jiang}}, \bibinfo
  {author} {\bibfnamefont {Z.}~\bibnamefont {Hao}}, \bibinfo {author}
  {\bibfnamefont {M.~C.}\ \bibnamefont {Martin}}, \bibinfo {author}
  {\bibfnamefont {P.}~\bibnamefont {Kim}}, \bibinfo {author} {\bibfnamefont
  {H.~L.}\ \bibnamefont {Stormer}}, \ and\ \bibinfo {author} {\bibfnamefont
  {D.~N.}\ \bibnamefont {Basov}},\ }\href@noop {} {\bibfield  {journal}
  {\bibinfo  {journal} {Nature Phys.}\ }\textbf {\bibinfo {volume} {4}},\
  \bibinfo {pages} {532} (\bibinfo {year} {2008})}\BibitemShut {NoStop}%
\bibitem [{\citenamefont {Kuzmenko}\ \emph {et~al.}(2008)\citenamefont
  {Kuzmenko}, \citenamefont {van Heumen}, \citenamefont {Carbone},\ and\
  \citenamefont {van~der Marel}}]{Kuzmenko:2008}%
  \BibitemOpen
  \bibfield  {author} {\bibinfo {author} {\bibfnamefont {A.~B.}\ \bibnamefont
  {Kuzmenko}}, \bibinfo {author} {\bibfnamefont {E.}~\bibnamefont {van
  Heumen}}, \bibinfo {author} {\bibfnamefont {F.}~\bibnamefont {Carbone}}, \
  and\ \bibinfo {author} {\bibfnamefont {D.}~\bibnamefont {van~der Marel}},\
  }\href@noop {} {\bibfield  {journal} {\bibinfo  {journal} {Phys. Rev. Lett.}\
  }\textbf {\bibinfo {volume} {100}},\ \bibinfo {pages} {117401} (\bibinfo
  {year} {2008})}\BibitemShut {NoStop}%
\bibitem [{\citenamefont {Nair}\ \emph {et~al.}(2008)\citenamefont {Nair},
  \citenamefont {Blake}, \citenamefont {Grigeronko}, \citenamefont {Novoselov},
  \citenamefont {Booth}, \citenamefont {Stauber}, \citenamefont {Peres},\ and\
  \citenamefont {Geim}}]{Nair:2008}%
  \BibitemOpen
  \bibfield  {author} {\bibinfo {author} {\bibfnamefont {R.~R.}\ \bibnamefont
  {Nair}}, \bibinfo {author} {\bibfnamefont {B.}~\bibnamefont {Blake}},
  \bibinfo {author} {\bibfnamefont {A.~N.}\ \bibnamefont {Grigeronko}},
  \bibinfo {author} {\bibfnamefont {K.~S.}\ \bibnamefont {Novoselov}}, \bibinfo
  {author} {\bibfnamefont {T.~J.}\ \bibnamefont {Booth}}, \bibinfo {author}
  {\bibfnamefont {T.}~\bibnamefont {Stauber}}, \bibinfo {author} {\bibfnamefont
  {N.~M.~R.}\ \bibnamefont {Peres}}, \ and\ \bibinfo {author} {\bibfnamefont
  {A.~K.}\ \bibnamefont {Geim}},\ }\href@noop {} {\bibfield  {journal}
  {\bibinfo  {journal} {Science}\ }\textbf {\bibinfo {volume} {320}},\ \bibinfo
  {pages} {1308} (\bibinfo {year} {2008})}\BibitemShut {NoStop}%
\bibitem [{\citenamefont {Mak}\ \emph {et~al.}(2008)\citenamefont {Mak},
  \citenamefont {Sfeir}, \citenamefont {Wu}, \citenamefont {Lui}, \citenamefont
  {Misewich},\ and\ \citenamefont {Heinz}}]{Mak:2008}%
  \BibitemOpen
  \bibfield  {author} {\bibinfo {author} {\bibfnamefont {K.~F.}\ \bibnamefont
  {Mak}}, \bibinfo {author} {\bibfnamefont {M.~Y.}\ \bibnamefont {Sfeir}},
  \bibinfo {author} {\bibfnamefont {Y.}~\bibnamefont {Wu}}, \bibinfo {author}
  {\bibfnamefont {C.~H.}\ \bibnamefont {Lui}}, \bibinfo {author} {\bibfnamefont
  {J.~A.}\ \bibnamefont {Misewich}}, \ and\ \bibinfo {author} {\bibfnamefont
  {T.~F.}\ \bibnamefont {Heinz}},\ }\href@noop {} {\bibfield  {journal}
  {\bibinfo  {journal} {Phys. Rev. Lett.}\ }\textbf {\bibinfo {volume} {101}},\
  \bibinfo {pages} {196405} (\bibinfo {year} {2008})}\BibitemShut {NoStop}%
\bibitem [{\citenamefont {Abergel}\ and\ \citenamefont
  {Fal'ko}(2007)}]{Abergel:2007}%
  \BibitemOpen
  \bibfield  {author} {\bibinfo {author} {\bibfnamefont {D.~S.~L.}\
  \bibnamefont {Abergel}}\ and\ \bibinfo {author} {\bibfnamefont {V.~I.}\
  \bibnamefont {Fal'ko}},\ }\href@noop {} {\bibfield  {journal} {\bibinfo
  {journal} {Phys. Rev. B}\ }\textbf {\bibinfo {volume} {75}},\ \bibinfo
  {pages} {155430} (\bibinfo {year} {2007})}\BibitemShut {NoStop}%
\bibitem [{\citenamefont {Nicol}\ and\ \citenamefont
  {Carbotte}(2008)}]{Nicol:2008}%
  \BibitemOpen
  \bibfield  {author} {\bibinfo {author} {\bibfnamefont {E.~J.}\ \bibnamefont
  {Nicol}}\ and\ \bibinfo {author} {\bibfnamefont {J.~P.}\ \bibnamefont
  {Carbotte}},\ }\href@noop {} {\bibfield  {journal} {\bibinfo  {journal}
  {Phys. Rev. B}\ }\textbf {\bibinfo {volume} {77}},\ \bibinfo {pages} {155409}
  (\bibinfo {year} {2008})}\BibitemShut {NoStop}%
\bibitem [{\citenamefont {Nilsson}\ \emph {et~al.}(2008)\citenamefont
  {Nilsson}, \citenamefont {{Castro Neto}}, \citenamefont {Guinea},\ and\
  \citenamefont {Peres}}]{Nilsson:2008}%
  \BibitemOpen
  \bibfield  {author} {\bibinfo {author} {\bibfnamefont {J.}~\bibnamefont
  {Nilsson}}, \bibinfo {author} {\bibfnamefont {A.~H.}\ \bibnamefont {{Castro
  Neto}}}, \bibinfo {author} {\bibfnamefont {F.}~\bibnamefont {Guinea}}, \ and\
  \bibinfo {author} {\bibfnamefont {N.~M.~R.}\ \bibnamefont {Peres}},\
  }\href@noop {} {\bibfield  {journal} {\bibinfo  {journal} {Phys. Rev. B}\
  }\textbf {\bibinfo {volume} {78}},\ \bibinfo {pages} {045405} (\bibinfo
  {year} {2008})}\BibitemShut {NoStop}%
\bibitem [{\citenamefont {Koshino}\ and\ \citenamefont
  {Ando}(2009)}]{Koshino:2009}%
  \BibitemOpen
  \bibfield  {author} {\bibinfo {author} {\bibfnamefont {M.}~\bibnamefont
  {Koshino}}\ and\ \bibinfo {author} {\bibfnamefont {T.}~\bibnamefont {Ando}},\
  }\href@noop {} {\bibfield  {journal} {\bibinfo  {journal} {Solid State
  Commun.}\ }\textbf {\bibinfo {volume} {149}},\ \bibinfo {pages} {1123}
  (\bibinfo {year} {2009})}\BibitemShut {NoStop}%
\bibitem [{\citenamefont {Li}\ \emph {et~al.}(2009)\citenamefont {Li},
  \citenamefont {Henriksen}, \citenamefont {Jiang}, \citenamefont {Hao},
  \citenamefont {Martin}, \citenamefont {Kim}, \citenamefont {Stormer},\ and\
  \citenamefont {Basov}}]{Li:2009}%
  \BibitemOpen
  \bibfield  {author} {\bibinfo {author} {\bibfnamefont {Z.~Q.}\ \bibnamefont
  {Li}}, \bibinfo {author} {\bibfnamefont {E.~A.}\ \bibnamefont {Henriksen}},
  \bibinfo {author} {\bibfnamefont {Z.}~\bibnamefont {Jiang}}, \bibinfo
  {author} {\bibfnamefont {Z.}~\bibnamefont {Hao}}, \bibinfo {author}
  {\bibfnamefont {M.~C.}\ \bibnamefont {Martin}}, \bibinfo {author}
  {\bibfnamefont {P.}~\bibnamefont {Kim}}, \bibinfo {author} {\bibfnamefont
  {H.~L.}\ \bibnamefont {Stormer}}, \ and\ \bibinfo {author} {\bibfnamefont
  {D.~N.}\ \bibnamefont {Basov}},\ }\href@noop {} {\bibfield  {journal}
  {\bibinfo  {journal} {Phys. Rev. Lett.}\ }\textbf {\bibinfo {volume} {102}},\
  \bibinfo {pages} {037403} (\bibinfo {year} {2009})}\BibitemShut {NoStop}%
\bibitem [{\citenamefont {Zhang}\ \emph {et~al.}(2008)\citenamefont {Zhang},
  \citenamefont {Li}, \citenamefont {Basov}, \citenamefont {Fogler},
  \citenamefont {Hao},\ and\ \citenamefont {Martin}}]{Zhang:2008}%
  \BibitemOpen
  \bibfield  {author} {\bibinfo {author} {\bibfnamefont {L.~M.}\ \bibnamefont
  {Zhang}}, \bibinfo {author} {\bibfnamefont {Z.~Q.}\ \bibnamefont {Li}},
  \bibinfo {author} {\bibfnamefont {D.~N.}\ \bibnamefont {Basov}}, \bibinfo
  {author} {\bibfnamefont {M.~M.}\ \bibnamefont {Fogler}}, \bibinfo {author}
  {\bibfnamefont {Z.}~\bibnamefont {Hao}}, \ and\ \bibinfo {author}
  {\bibfnamefont {M.~C.}\ \bibnamefont {Martin}},\ }\href@noop {} {\bibfield
  {journal} {\bibinfo  {journal} {Phys. Rev. B}\ }\textbf {\bibinfo {volume}
  {78}},\ \bibinfo {pages} {235408} (\bibinfo {year} {2008})}\BibitemShut
  {NoStop}%
\bibitem [{\citenamefont {Kuzmenko}\ \emph {et~al.}(2009)\citenamefont
  {Kuzmenko}, \citenamefont {van Heumen}, \citenamefont {van~der Marel},
  \citenamefont {Lerch}, \citenamefont {Blake}, \citenamefont {Novoselov},\
  and\ \citenamefont {Geim}}]{Kuzmenko:2009}%
  \BibitemOpen
  \bibfield  {author} {\bibinfo {author} {\bibfnamefont {A.~B.}\ \bibnamefont
  {Kuzmenko}}, \bibinfo {author} {\bibfnamefont {E.}~\bibnamefont {van
  Heumen}}, \bibinfo {author} {\bibfnamefont {D.}~\bibnamefont {van~der
  Marel}}, \bibinfo {author} {\bibfnamefont {P.}~\bibnamefont {Lerch}},
  \bibinfo {author} {\bibfnamefont {P.}~\bibnamefont {Blake}}, \bibinfo
  {author} {\bibfnamefont {K.~S.}\ \bibnamefont {Novoselov}}, \ and\ \bibinfo
  {author} {\bibfnamefont {A.~K.}\ \bibnamefont {Geim}},\ }\href@noop {}
  {\bibfield  {journal} {\bibinfo  {journal} {Phys. Rev. B}\ }\textbf {\bibinfo
  {volume} {79}},\ \bibinfo {pages} {115441} (\bibinfo {year}
  {2009})}\BibitemShut {NoStop}%
\bibitem [{\citenamefont {Orlita}\ and\ \citenamefont
  {Potemski}(2010)}]{Orlita:2010}%
  \BibitemOpen
  \bibfield  {author} {\bibinfo {author} {\bibfnamefont {M.}~\bibnamefont
  {Orlita}}\ and\ \bibinfo {author} {\bibfnamefont {M.}~\bibnamefont
  {Potemski}},\ }\href@noop {} {\bibfield  {journal} {\bibinfo  {journal}
  {Semicond. Sci. Technol.}\ }\textbf {\bibinfo {volume} {25}},\ \bibinfo
  {pages} {063001} (\bibinfo {year} {2010})}\BibitemShut {NoStop}%
\bibitem [{\citenamefont {Xu}\ \emph {et~al.}(2010)\citenamefont {Xu},
  \citenamefont {Li},\ and\ \citenamefont {Dong}}]{Xu:2010}%
  \BibitemOpen
  \bibfield  {author} {\bibinfo {author} {\bibfnamefont {Y.}~\bibnamefont
  {Xu}}, \bibinfo {author} {\bibfnamefont {X.}~\bibnamefont {Li}}, \ and\
  \bibinfo {author} {\bibfnamefont {J.}~\bibnamefont {Dong}},\ }\href@noop {}
  {\bibfield  {journal} {\bibinfo  {journal} {Nanotechnology}\ }\textbf
  {\bibinfo {volume} {21}},\ \bibinfo {pages} {065711} (\bibinfo {year}
  {2010})}\BibitemShut {NoStop}%
\bibitem [{\citenamefont {Prada}\ \emph {et~al.}(2011)\citenamefont {Prada},
  \citenamefont {San-Jose}, \citenamefont {Brey},\ and\ \citenamefont
  {Fertig}}]{Prada:2011}%
  \BibitemOpen
  \bibfield  {author} {\bibinfo {author} {\bibfnamefont {E.}~\bibnamefont
  {Prada}}, \bibinfo {author} {\bibfnamefont {P.}~\bibnamefont {San-Jose}},
  \bibinfo {author} {\bibfnamefont {L.}~\bibnamefont {Brey}}, \ and\ \bibinfo
  {author} {\bibfnamefont {H.~A.}\ \bibnamefont {Fertig}},\ }\href@noop {}
  {\bibfield  {journal} {\bibinfo  {journal} {Solid State Commun.}\ }\textbf
  {\bibinfo {volume} {151}},\ \bibinfo {pages} {1075} (\bibinfo {year}
  {2011})}\BibitemShut {NoStop}%
\bibitem [{\citenamefont {Stauber}\ and\ \citenamefont
  {Peres}(2008)}]{Stauber:2008a}%
  \BibitemOpen
  \bibfield  {author} {\bibinfo {author} {\bibfnamefont {T.}~\bibnamefont
  {Stauber}}\ and\ \bibinfo {author} {\bibfnamefont {N.~M.~R.}\ \bibnamefont
  {Peres}},\ }\href@noop {} {\bibfield  {journal} {\bibinfo  {journal} {J.
  Phys. Condens. Matter}\ }\textbf {\bibinfo {volume} {20}},\ \bibinfo {pages}
  {055002} (\bibinfo {year} {2008})}\BibitemShut {NoStop}%
\bibitem [{\citenamefont {Nicol}\ and\ \citenamefont
  {Carbotte}(2009)}]{Nicol:2009}%
  \BibitemOpen
  \bibfield  {author} {\bibinfo {author} {\bibfnamefont {E.~J.}\ \bibnamefont
  {Nicol}}\ and\ \bibinfo {author} {\bibfnamefont {J.~P.}\ \bibnamefont
  {Carbotte}},\ }\href@noop {} {\bibfield  {journal} {\bibinfo  {journal}
  {Phys. Rev. B}\ }\textbf {\bibinfo {volume} {80}},\ \bibinfo {pages}
  {081415(R)} (\bibinfo {year} {2009})}\BibitemShut {NoStop}%
\bibitem [{\citenamefont {Carbotte}\ \emph {et~al.}(2010)\citenamefont
  {Carbotte}, \citenamefont {Nicol},\ and\ \citenamefont
  {Sharapov}}]{Carbotte:2010}%
  \BibitemOpen
  \bibfield  {author} {\bibinfo {author} {\bibfnamefont {J.~P.}\ \bibnamefont
  {Carbotte}}, \bibinfo {author} {\bibfnamefont {E.~J.}\ \bibnamefont {Nicol}},
  \ and\ \bibinfo {author} {\bibfnamefont {S.~G.}\ \bibnamefont {Sharapov}},\
  }\href@noop {} {\bibfield  {journal} {\bibinfo  {journal} {Phys. Rev. B}\
  }\textbf {\bibinfo {volume} {81}},\ \bibinfo {pages} {045419} (\bibinfo
  {year} {2010})}\BibitemShut {NoStop}%
\bibitem [{\citenamefont {Pound}\ \emph
  {et~al.}(2011{\natexlab{a}})\citenamefont {Pound}, \citenamefont {Carbotte},\
  and\ \citenamefont {Nicol}}]{Pound:2011a}%
  \BibitemOpen
  \bibfield  {author} {\bibinfo {author} {\bibfnamefont {A.}~\bibnamefont
  {Pound}}, \bibinfo {author} {\bibfnamefont {J.~P.}\ \bibnamefont {Carbotte}},
  \ and\ \bibinfo {author} {\bibfnamefont {E.~J.}\ \bibnamefont {Nicol}},\
  }\href@noop {} {\bibfield  {journal} {\bibinfo  {journal} {Europhys. Lett.}\
  }\textbf {\bibinfo {volume} {94}},\ \bibinfo {pages} {57006} (\bibinfo {year}
  {2011}{\natexlab{a}})}\BibitemShut {NoStop}%
\bibitem [{\citenamefont {Pound}\ \emph
  {et~al.}(2011{\natexlab{b}})\citenamefont {Pound}, \citenamefont {Carbotte},\
  and\ \citenamefont {Nicol}}]{Pound:2011b}%
  \BibitemOpen
  \bibfield  {author} {\bibinfo {author} {\bibfnamefont {A.}~\bibnamefont
  {Pound}}, \bibinfo {author} {\bibfnamefont {J.~P.}\ \bibnamefont {Carbotte}},
  \ and\ \bibinfo {author} {\bibfnamefont {E.~J.}\ \bibnamefont {Nicol}},\
  }\href@noop {} {\bibfield  {journal} {\bibinfo  {journal} {Phys. Rev. B}\
  }\textbf {\bibinfo {volume} {84}},\ \bibinfo {pages} {085125} (\bibinfo
  {year} {2011}{\natexlab{b}})}\BibitemShut {NoStop}%
\bibitem [{\citenamefont {Pound}\ \emph {et~al.}(2012)\citenamefont {Pound},
  \citenamefont {Carbotte},\ and\ \citenamefont {Nicol}}]{Pound:2012}%
  \BibitemOpen
  \bibfield  {author} {\bibinfo {author} {\bibfnamefont {A.}~\bibnamefont
  {Pound}}, \bibinfo {author} {\bibfnamefont {J.~P.}\ \bibnamefont {Carbotte}},
  \ and\ \bibinfo {author} {\bibfnamefont {E.~J.}\ \bibnamefont {Nicol}},\
  }\href@noop {} {\bibfield  {journal} {\bibinfo  {journal} {Phys. Rev. B}\
  }\textbf {\bibinfo {volume} {85}},\ \bibinfo {pages} {125422} (\bibinfo
  {year} {2012})}\BibitemShut {NoStop}%
\bibitem [{\citenamefont {LeBlanc}\ \emph {et~al.}(2011)\citenamefont
  {LeBlanc}, \citenamefont {Carbotte},\ and\ \citenamefont
  {Nicol}}]{LeBlanc:2011}%
  \BibitemOpen
  \bibfield  {author} {\bibinfo {author} {\bibfnamefont {J.~P.~F.}\
  \bibnamefont {LeBlanc}}, \bibinfo {author} {\bibfnamefont {J.~P.}\
  \bibnamefont {Carbotte}}, \ and\ \bibinfo {author} {\bibfnamefont {E.~J.}\
  \bibnamefont {Nicol}},\ }\href@noop {} {\bibfield  {journal} {\bibinfo
  {journal} {Phys. Rev. B}\ }\textbf {\bibinfo {volume} {84}},\ \bibinfo
  {pages} {165448} (\bibinfo {year} {2011})}\BibitemShut {NoStop}%
\bibitem [{\citenamefont {Carbotte}\ \emph {et~al.}(2012)\citenamefont
  {Carbotte}, \citenamefont {LeBlanc},\ and\ \citenamefont
  {Nicol}}]{Carbotte:2012}%
  \BibitemOpen
  \bibfield  {author} {\bibinfo {author} {\bibfnamefont {J.~P.}\ \bibnamefont
  {Carbotte}}, \bibinfo {author} {\bibfnamefont {J.~P.~F.}\ \bibnamefont
  {LeBlanc}}, \ and\ \bibinfo {author} {\bibfnamefont {E.~J.}\ \bibnamefont
  {Nicol}},\ }\href@noop {} {\bibfield  {journal} {\bibinfo  {journal} {Phys.
  Rev. B}\ }\textbf {\bibinfo {volume} {85}},\ \bibinfo {pages} {201411(R)}
  (\bibinfo {year} {2012})}\BibitemShut {NoStop}%
\bibitem [{\citenamefont {Principi}\ \emph {et~al.}(2012)\citenamefont
  {Principi}, \citenamefont {Polini}, \citenamefont {Asgari},\ and\
  \citenamefont {MacDonald}}]{Principi:2012}%
  \BibitemOpen
  \bibfield  {author} {\bibinfo {author} {\bibfnamefont {A.}~\bibnamefont
  {Principi}}, \bibinfo {author} {\bibfnamefont {M.}~\bibnamefont {Polini}},
  \bibinfo {author} {\bibfnamefont {R.}~\bibnamefont {Asgari}}, \ and\ \bibinfo
  {author} {\bibfnamefont {A.~H.}\ \bibnamefont {MacDonald}},\ }\href@noop {}
  {} (\bibinfo {year} {2012}),\ \Eprint {http://arxiv.org/abs/arXiv:1111.3822}
  {arXiv:1111.3822} \BibitemShut {NoStop}%
\bibitem [{\citenamefont {Mahan}(1990)}]{Mahan:1990}%
  \BibitemOpen
  \bibfield  {author} {\bibinfo {author} {\bibfnamefont {G.~D.}\ \bibnamefont
  {Mahan}},\ }\href@noop {} {\emph {\bibinfo {title} {Many Particle Physics}}}\
  (\bibinfo  {publisher} {Plenum},\ \bibinfo {address} {New York},\ \bibinfo
  {year} {1990})\BibitemShut {NoStop}%
\bibitem [{\citenamefont {Lobato}\ and\ \citenamefont
  {Partoens}(2011)}]{Lobato:2011}%
  \BibitemOpen
  \bibfield  {author} {\bibinfo {author} {\bibfnamefont {I.}~\bibnamefont
  {Lobato}}\ and\ \bibinfo {author} {\bibfnamefont {B.}~\bibnamefont
  {Partoens}},\ }\href@noop {} {\bibfield  {journal} {\bibinfo  {journal}
  {Phys. Rev. B}\ }\textbf {\bibinfo {volume} {83}},\ \bibinfo {pages} {165429}
  (\bibinfo {year} {2011})}\BibitemShut {NoStop}%
\bibitem [{\citenamefont {Charlier}\ \emph {et~al.}(1992)\citenamefont
  {Charlier}, \citenamefont {Michenaud},\ and\ \citenamefont
  {Gonze}}]{Charlier:1992}%
  \BibitemOpen
  \bibfield  {author} {\bibinfo {author} {\bibfnamefont {J.-C.}\ \bibnamefont
  {Charlier}}, \bibinfo {author} {\bibfnamefont {J.-P.}\ \bibnamefont
  {Michenaud}}, \ and\ \bibinfo {author} {\bibfnamefont {X.}~\bibnamefont
  {Gonze}},\ }\href@noop {} {\bibfield  {journal} {\bibinfo  {journal} {Phys.
  Rev. B}\ }\textbf {\bibinfo {volume} {46}},\ \bibinfo {pages} {4531}
  (\bibinfo {year} {1992})}\BibitemShut {NoStop}%
\bibitem [{\citenamefont {McCann}(2006)}]{McCann:2006}%
  \BibitemOpen
  \bibfield  {author} {\bibinfo {author} {\bibfnamefont {E.}~\bibnamefont
  {McCann}},\ }\href@noop {} {\bibfield  {journal} {\bibinfo  {journal} {Phys.
  Rev. B}\ }\textbf {\bibinfo {volume} {74}},\ \bibinfo {pages} {161403(R)}
  (\bibinfo {year} {2006})}\BibitemShut {NoStop}%
\bibitem [{\citenamefont {Ando}(2011)}]{Ando:2011}%
  \BibitemOpen
  \bibfield  {author} {\bibinfo {author} {\bibfnamefont {T.}~\bibnamefont
  {Ando}},\ }\href@noop {} {\bibfield  {journal} {\bibinfo  {journal} {J.
  Phys.: Conf. Ser.}\ }\textbf {\bibinfo {volume} {302}},\ \bibinfo {pages}
  {012015} (\bibinfo {year} {2011})}\BibitemShut {NoStop}%
\bibitem [{\citenamefont {Sensale-Rodriguez}\ \emph {et~al.}(2012)\citenamefont
  {Sensale-Rodriguez}, \citenamefont {Yan}, \citenamefont {Kelly},
  \citenamefont {Fang},\ and\ \citenamefont {Tahy}}]{Sensale:2012}%
  \BibitemOpen
  \bibfield  {author} {\bibinfo {author} {\bibfnamefont {B.}~\bibnamefont
  {Sensale-Rodriguez}}, \bibinfo {author} {\bibfnamefont {R.}~\bibnamefont
  {Yan}}, \bibinfo {author} {\bibfnamefont {M.~M.}\ \bibnamefont {Kelly}},
  \bibinfo {author} {\bibfnamefont {T.}~\bibnamefont {Fang}}, \ and\ \bibinfo
  {author} {\bibfnamefont {K.}~\bibnamefont {Tahy}},\ }\href {\doibase
  10.1038/ncomms1797} {\bibfield  {journal} {\bibinfo  {journal} {Nature
  Commun.}\ }\textbf {\bibinfo {volume} {3}},\ \bibinfo {pages} {780} (\bibinfo
  {year} {2012})}\BibitemShut {NoStop}%
\bibitem [{\citenamefont {Ando}\ and\ \citenamefont
  {Koshino}(2009)}]{Ando:2009}%
  \BibitemOpen
  \bibfield  {author} {\bibinfo {author} {\bibfnamefont {T.}~\bibnamefont
  {Ando}}\ and\ \bibinfo {author} {\bibfnamefont {M.}~\bibnamefont {Koshino}},\
  }\href@noop {} {\bibfield  {journal} {\bibinfo  {journal} {J. Phys. Soc.
  Jpn.}\ }\textbf {\bibinfo {volume} {78}},\ \bibinfo {pages} {104716}
  (\bibinfo {year} {2009})}\BibitemShut {NoStop}%
\bibitem [{\citenamefont {Kane}\ and\ \citenamefont {Mele}(2005)}]{Kane:2005}%
  \BibitemOpen
  \bibfield  {author} {\bibinfo {author} {\bibfnamefont {C.~L.}\ \bibnamefont
  {Kane}}\ and\ \bibinfo {author} {\bibfnamefont {E.~J.}\ \bibnamefont
  {Mele}},\ }\href@noop {} {\bibfield  {journal} {\bibinfo  {journal} {Phys.
  Rev. Lett.}\ }\textbf {\bibinfo {volume} {95}},\ \bibinfo {pages} {226801}
  (\bibinfo {year} {2005})}\BibitemShut {NoStop}%
\bibitem [{\citenamefont {Liu}\ \emph {et~al.}(2011)\citenamefont {Liu},
  \citenamefont {Feng},\ and\ \citenamefont {Yao}}]{Liu:2011}%
  \BibitemOpen
  \bibfield  {author} {\bibinfo {author} {\bibfnamefont {C.-C.}\ \bibnamefont
  {Liu}}, \bibinfo {author} {\bibfnamefont {W.}~\bibnamefont {Feng}}, \ and\
  \bibinfo {author} {\bibfnamefont {Y.}~\bibnamefont {Yao}},\ }\href {\doibase
  10.1103/PhysRevLett.107.076802} {\bibfield  {journal} {\bibinfo  {journal}
  {Phys. Rev. Lett.}\ }\textbf {\bibinfo {volume} {107}},\ \bibinfo {pages}
  {076802} (\bibinfo {year} {2011})}\BibitemShut {NoStop}%
\bibitem [{\citenamefont {Drummond}\ \emph {et~al.}(2012)\citenamefont
  {Drummond}, \citenamefont {Z\'olyomi},\ and\ \citenamefont
  {Fal'ko}}]{Drummond:2012}%
  \BibitemOpen
  \bibfield  {author} {\bibinfo {author} {\bibfnamefont {N.~D.}\ \bibnamefont
  {Drummond}}, \bibinfo {author} {\bibfnamefont {V.}~\bibnamefont {Z\'olyomi}},
  \ and\ \bibinfo {author} {\bibfnamefont {V.~I.}\ \bibnamefont {Fal'ko}},\
  }\href {\doibase 10.1103/PhysRevB.85.075423} {\bibfield  {journal} {\bibinfo
  {journal} {Phys. Rev. B}\ }\textbf {\bibinfo {volume} {85}},\ \bibinfo
  {pages} {075423} (\bibinfo {year} {2012})}\BibitemShut {NoStop}%
\bibitem [{\citenamefont {Sato}\ \emph {et~al.}(2011)\citenamefont {Sato},
  \citenamefont {Segawa}, \citenamefont {Kosaka}, \citenamefont {Souma},
  \citenamefont {Nakayama}, \citenamefont {Eto}, \citenamefont {Minami},
  \citenamefont {Ando},\ and\ \citenamefont {Takahashi}}]{Sato:2011}%
  \BibitemOpen
  \bibfield  {author} {\bibinfo {author} {\bibfnamefont {T.}~\bibnamefont
  {Sato}}, \bibinfo {author} {\bibfnamefont {K.}~\bibnamefont {Segawa}},
  \bibinfo {author} {\bibfnamefont {K.}~\bibnamefont {Kosaka}}, \bibinfo
  {author} {\bibfnamefont {S.}~\bibnamefont {Souma}}, \bibinfo {author}
  {\bibfnamefont {K.}~\bibnamefont {Nakayama}}, \bibinfo {author}
  {\bibfnamefont {K.}~\bibnamefont {Eto}}, \bibinfo {author} {\bibfnamefont
  {T.}~\bibnamefont {Minami}}, \bibinfo {author} {\bibfnamefont
  {Y.}~\bibnamefont {Ando}}, \ and\ \bibinfo {author} {\bibfnamefont
  {T.}~\bibnamefont {Takahashi}},\ }\href@noop {} {\bibfield  {journal}
  {\bibinfo  {journal} {Nature Phys.}\ }\textbf {\bibinfo {volume} {7}},\
  \bibinfo {pages} {840} (\bibinfo {year} {2011})}\BibitemShut {NoStop}%
\bibitem [{\citenamefont {Xiao}\ \emph {et~al.}(2012)\citenamefont {Xiao},
  \citenamefont {Liu}, \citenamefont {Feng}, \citenamefont {Xu},\ and\
  \citenamefont {Yao}}]{Xiao:2012}%
  \BibitemOpen
  \bibfield  {author} {\bibinfo {author} {\bibfnamefont {D.}~\bibnamefont
  {Xiao}}, \bibinfo {author} {\bibfnamefont {G.-B.}\ \bibnamefont {Liu}},
  \bibinfo {author} {\bibfnamefont {W.}~\bibnamefont {Feng}}, \bibinfo {author}
  {\bibfnamefont {X.}~\bibnamefont {Xu}}, \ and\ \bibinfo {author}
  {\bibfnamefont {W.}~\bibnamefont {Yao}},\ }\href {\doibase
  10.1103/PhysRevLett.108.196802} {\bibfield  {journal} {\bibinfo  {journal}
  {Phys. Rev. Lett.}\ }\textbf {\bibinfo {volume} {108}},\ \bibinfo {pages}
  {196802} (\bibinfo {year} {2012})}\BibitemShut {NoStop}%
\bibitem [{\citenamefont {Tse}\ and\ \citenamefont
  {MacDonald}(2010)}]{Tse:2010}%
  \BibitemOpen
  \bibfield  {author} {\bibinfo {author} {\bibfnamefont {W.-K.}\ \bibnamefont
  {Tse}}\ and\ \bibinfo {author} {\bibfnamefont {A.~H.}\ \bibnamefont
  {MacDonald}},\ }\href@noop {} {\bibfield  {journal} {\bibinfo  {journal}
  {Phys. Rev. Lett.}\ }\textbf {\bibinfo {volume} {105}},\ \bibinfo {pages}
  {057401} (\bibinfo {year} {2010})}\BibitemShut {NoStop}%
\end{thebibliography}%

\end{document}